\documentclass[aps,pra,showpacs,superscriptaddress,amssymb,twocolumn]{revtex4-1}
\pdfoutput=1
\usepackage{eqnarray,amsmath,braket,physics}
\usepackage{natbib}
\usepackage{graphicx}
\usepackage{enumitem}
\usepackage{appendix}
\usepackage{color}
\usepackage[export]{adjustbox}
\usepackage{epstopdf} 
\DeclareGraphicsExtensions{.pdf,.eps,.png,.jpg,.mps}
\usepackage[colorlinks=true,citecolor=blue,urlcolor=blue,linkcolor=blue]{hyperref}
\usepackage[position=top,singlelinecheck=off,justification=raggedright,caption=false]{subfig}
\usepackage{bm,amsmath}
\usepackage{dcolumn}
\usepackage{multirow}
\usepackage{hyperref}
\usepackage{float}
\newcommand\iinteg[4]{(#1 \ #2 \ | \ #3 \ #4)}

\def\BEq{\begin{equation}}
\def\EEq{\end{equation}}
\def\BEqA{\begin{eqnarray}}
\def\EEqA{\end{eqnarray}}
\def\BW{\begin{widetext}}
\def\EW{\end{widetext}}
\usepackage{array}

\begin{document}
\title{Strong electron-electron interactions in Si/SiGe quantum dots}

\author{H.\ Ekmel Ercan}
\affiliation{Department of Physics, University of Wisconsin-Madison, Madison, Wisconsin 53706, USA}
\author{S.\ N.\ Coppersmith}
\affiliation{Department of Physics, University of Wisconsin-Madison, Madison, Wisconsin 53706, USA}
\affiliation{School of Physics, The University of New South Wales
Sydney, NSW 2052, Australia}
\author{Mark Friesen}
\affiliation{Department of Physics, University of Wisconsin-Madison, Madison, Wisconsin 53706, USA}

\date{\today}

\begin{abstract}
Interactions between electrons can strongly affect the shape and functionality of multi-electron quantum dots.
The resulting charge distributions can be localized, as in the case of Wigner molecules, with consequences for the energy spectrum and tunneling to states outside the dot.
The situation is even more complicated for silicon dots, due to the interplay between valley, orbital, and interaction energy scales.
Here, we study  two-electron wavefunctions in electrostatically confined quantum dots formed in a SiGe/Si/SiGe quantum well at zero magnetic field, using a combination of tight-binding and full-configuration-interaction (FCI) methods, and taking into account atomic-scale disorder at the quantum well interface.
We model dots based on recent qubit experiments, which straddle the boundary between strongly interacting and weakly interacting systems, and display a rich and diverse range of behaviors.
Our calculations show that strong electron-electron interactions, induced by weak confinement, can significantly suppress the low-lying, singlet-triplet (ST) excitation energy.
However, when the valley-orbit interactions caused by interfacial disorder are weak, the ST splitting can approach its noninteracting value, even when the electron-electron interactions are strong and Wigner-molecule behavior is observed.
These results have important implications for the rational design and fabrication of quantum dot qubits with predictable properties.

\end{abstract}    

\maketitle

\section{Introduction}

Silicon quantum dots are of great current interest because of their desirable properties as qubits~\cite{Loss:1998p120,Zwanenburg:2013p961,Zhang:2018p32,Borselli:2011p063109,Kim:2014p70,Veldhorst:2015p410,Watson:2018p633,Andrews:2019p05004,Harvey:2019p217702,Borjans:2020p195}, including the high natural abundance of spin-0 nuclear isotopes, and the possibility of leveraging the infrastructure of the semiconductor industry for scale-up.
Device uniformity plays a key role in scale-up, and recent, new fabrication schemes have demonstrated improved control over geometry-influenced device parameters, including dot confinement energies and tunnel barriers~\cite{Angus:2007p845,Zajac:2015p223507,Zajac:2016p054013, Mi:2018p599,Zajac:2018p439,Neyens:2019p08216}.
However, new physics enters the picture for dots and qubits with more than one electron~\cite{Kodera:2009p22043,Kim:2014p70,Chen:2017p35408,Harvey:2017p1029,Leon:2020p797,Yang:2020p350,Petit:2020p355}, making it more difficult to achieve uniformity.
Electron-electron (e-e) interactions present a particular challenge, both from control and design perspectives, since they are so sensitive to many factors.
A  comprehensive and quantitative theoretical understanding of e-e interactions and the contributing physics is therefore highly desirable.

Strong e-e interactions occur in dots with low electron densities, giving rise to localization of electrons in the form of Wigner molecules~\cite{Bryant:1987p1140,Yannouleas:1999p5325,Egger:1999p3320,Filinov:2001p3851,Reusch:2001p113313}, which are finite-size analogs of Wigner crystals~\cite{Wigner:1934,Tanatar:1989}. 
Experimental signatures of Wigner molecules include the suppression of multi-electron excitation energies, which has been observed in GaAs~\cite{Singha:2010p246802} and Ga[Al]As~\cite{Ellenberger:2006p126806} dots and carbon nanotubes~\cite{Pecker:2013p576}. 
A detailed understanding of these charge distributions is particularly important for qubit applications, because of the need to precisely control the qubit energy splittings and tunnel couplings. 
Here we show that e-e interactions are particularly subtle for dots formed in SiGe/Si/SiGe quantum wells, because of the near degeneracy of the low-lying conduction-band valleys~\cite{Boykin:2004p115,Goswami:2007p41,Chutia:2008p193311}.
These many-body effects depend sensitively on the coupling between valley and orbital degrees of freedom, which is strongly affected by disorder at the Si/SiGe quantum well interface~\cite{Friesen:2006p202106,Friesen:2007p115318}, 

The importance of e-e interactions can be estimated from the Wigner ratio $R_W=E_\text{ee}/E_\text{orb}$, relating the Coulomb energy, $E_\text{ee}$, to the single-electron orbital splitting, $E_\text{orb}$.
Here, we assume a simple parabolic confinement potential, $m_t\omega^2x^2/2$, where $m_t$ is the effective mass in the plane of the quantum well.
The corresponding orbital energy is given by $E_\text{orb}=\hbar\omega$, the dot radius is given by $a_{\mathrm{0}}=\sqrt{\hbar / m_t \omega}$, and the characteristic Coulomb energy is given by $E_\text{ee}=e^2/4\pi\epsilon a_0$. 
As expected, the Coulomb energy increases with confinement strength; however, the orbital energy increases faster, resulting in e-e interactions that increase with dot size.
For two-dimensional (2D) circular dots, it has been theoretically estimated that Wigner molecules should emerge when the Wigner parameter $R_W\gtrsim 1.5$~\cite{Yannouleas:1999p5325}.
Such behavior has been experimentally confirmed in GaAs dots~\cite{Ellenberger:2006p126806}.
For Si dots, the $R_W= 1.5$ criterion corresponds to $\hbar \omega = 18 \ \mathrm{meV}$; however, orbital energies are typically of order $\hbar\omega\sim 1 \ \mathrm{meV}$, which corresponds to $R_W=6$.
We therefore expect Si qubits to be in the strongly interacting regime.

Valley physics plays a critical role with regards to e-e interactions.
For SiGe/Si/SiGe quantum wells, the presence of a sharp interface lifts the two-fold degeneracy of the conduction-band valleys.
However, this valley splitting is suppressed well below the orbital splitting, due to the unavoidable interfacial disorder present in realistic devices~\cite{Friesen:2007p115318,Friesen:2010p115324}.
The valley ratio $E_\text{ee}/E_\text{val}$ is therefore even larger than $R_W=E_\text{ee}/E_\text{orb}$.
While several previous calculations explore e-e interactions in silicon dots~\cite{Hada:2003p155322,Jiang:2013p85311,Miserev:2019p205129,Wang:2010p235326,Liu:2012p45311}, including Wigner molecule effects~\cite{Shaji:2008p540}, none includes the realistic disorder profiles that are key for understanding the valley physics.

In this paper we investigate the combined effects of e-e interactions and valley-orbit couplings (VOC) on the properties of the electronic states of a single quantum dot containing two electrons.
We focus on the ground state (a singlet), and the first two excited  triplet states.
Such single-dot ST energy splittings are key for many quantum computing applications.
This obviously includes ST qubits, particularly when tuned towards the readout regime~\cite{Levy:2002p1446,Petta:2005p2180}. 
For the quantum hybrid qubit, the qubit energy splitting is largely determined by the ST splitting in the dot containing two electrons~\cite{Shi:2012p140503,Koh:2012p250503,Kim:2014p70,Ferraro:2014p1155,Kim:2015p15004,Cao:2016p086801,Thorgrimsson:2017p32,Ferraro:2018p1800040,Ferraro:2019p035310,Corrigan:2020preprint,Wonjin:2021preprint}. 
Large ST splittings are also key requirements for readout methods based on Pauli spin blockade~\cite{Gaudreau:2011p54,Maune:2012p344,Medford:2013p050501,Li:2018p7}. 

Nonperturbative techniques are needed to treat strong e-e interactions, particularly in the Wigner molecule regime, where two-electron wavefunctions have very different charge distributions than the low-lying single-electron eigenstates.
Here, we use the full configuration interaction (FCI) method, in which multi-electron Hamiltonians are expanded in large sets of ``configurations," which are in turn constructed from symmetrized products of single-electron eigenstates~\cite{Szabo:1996p91861}.
Special techniques are also required to treat the VOC that arises due to atomic-scale disorder at the Si/SiGe quantum-well interface.
(Here, we use the terminology ``valley-orbit coupling" to describe the mixing of valley and orbital states that would otherwise represent good quantum numbers in the absence of interfacial disorder.
This is in contrast with some other authors that use the term interchangeably with ``valley splitting.")
Here, we employ a minimal tight-binding (TB) model~\cite{Boykin:2004p115,Boykin:2004p165325,Friesen:2010p115324,Abadillo-Uriel:2018p165438} to compute the single-electron basis states that are incorporated into the FCI calculations.
This powerful and unique combination of methods enables us to explore the diverse and unexpected behavior observed in Si qubit experiments~\cite{Corrigan:2020preprint,Dodson:2021}.

We explore the combined effects of e-e interactions and VOC in stages.
First, we effectively turn off the VOC by considering a perfectly flat quantum well interface, with no atomic-scale disorder.
In this case, ``valley" and ``orbital" are good quantum numbers for the single-electron eigenstates, and these symmetries are also conferred upon the two-electron wavefunctions. 
The resulting two-electron ground state is a singlet, while the lowest excited state is a triplet.
There are two distinct types of low-energy triplets, which play a key role in this work.
We refer to the first type as a ``valley triplet," because it involves both ground and excited valley states.
We call the second type an ``orbital triplet," because it includes ground and excited orbital states, but no excited valley states.

The main results of the present calculations are summarized as follows.
A key experimental parameter for tuning e-e interactions is the orbital (i.e., lateral) confinement potential; we adopt this as the main tuning parameter in our simulations. 
For weak confinement, the orbital triplet has the lower energy, while for strong confinement, the valley triplet has the lower energy.
Although the charge distributions of all two-electron wavefunctions are affected by e-e interactions, and may even form Wigner molecules for weak confinement, it is interesting that only the orbital ST splitting is found to depend on the confinement strength in the regime relevant to experiments on Si/SiGe quantum dots.
The valley ST splitting is protected by the valley symmetry, and remains essentially unaffected by the confinement, with a value nearly identical to the valley splitting.
As a result, the low-energy excited state switches from an orbital triplet (for weak confinement) to a valley triplet (for strong confinement).
This crossover occurs in a regime that is typically relevant for Si/SiGe quantum dot qubit experiments.

\begin{table}
\caption{Some symbols and abbreviations used in this work.}
\begin{tabular}{l c c c c c c}
\toprule
 Symbol &&&&& Description \\
 \colrule
 S &&&&& Low-energy singlet \\
 $\mathrm{T_{val}}$ &&&&& Low-energy valley triplet \\
 $\mathrm{T_{orb}}$ &&&&& Low-energy orbital triplet \\
 ST &&&&& Single-triplet (qubit)\\
 FCI &&&&& Full configuration interaction \\
 TB &&&&& Tight binding\\
 e-e &&&&& Electron-electron (interactions) \\
 FFT &&&&& Fast Fourier transform \\
 VOC &&&&& Valley-orbit coupling \\
 $E_{\mathrm{val}}$ &&&&& Valley splitting \\
 $E_{\mathrm{orb}}$ &&&&& Orbital splitting \\
 $E_{\mathrm{S}}$ &&&&& Singlet energy \\
 $E_{\mathrm{T_{val}}}$ &&&&& Valley-triplet energy \\
 $E_{\mathrm{T_{orb}}}$ &&&&& Orbital-triplet energy \\
 $E_{\mathrm{ST_{val}}}$ &&&&& $E_{\mathrm{T_{val}}}-E_{\mathrm{S}}$ \\
 $E_{\mathrm{ST_{orb}}}$ &&&&& $E_{\mathrm{T_{orb}}}-E_{\mathrm{S}}$ \\
 $E_{\mathrm{ST}}$ &&&&& min$(E_{\mathrm{ST_{val}}},E_{\mathrm{ST_{orb}}})$ \\
 $E_{\mathrm{ee}}$ &&&&& Coulomb-interaction energy \\
 $D$ &&&&& Dot diameter \\
 $W$ &&&&& Step separation \\
 $x_s$ &&&&& Step position \\
 $d=|x_s|$ &&&&& Distance from dot to nearest step \\
\botrule
\end{tabular}
\label{tab:symbol_list}
\end{table}

Next, we introduce atomic-scale steps at the quantum-well interface, resulting in VOC that hybridizes the valley and orbital states, and suppresses the valley splitting.
Typically, we want a VOC that is fairly weak, so that it yields ST splittings large enough to be appropriate for qubit applications. 
In this regime, ``valley" is no longer a good quantum number; however, the triplet states remain approximately valley-like or orbital-like. The different triplet states hybridize, so that the valley ST splitting is also suppressed -- in some cases quite significantly. 

An important take-home message from our analysis is that, in the strongly confined regime (where the valley triplet has lower energy than the orbital triplet), when large ST splittings are observed in experiments, they provide evidence of weak interfacial disorder; the measured ST splitting is then nearly equal to the valley splitting.
In this regime, the ST splitting is robust against small changes in the device tuning.
However, reducing the confinement strength lowers the energy of the orbital triplet relative to the valley triplet and also increases the importance of e-e interactions, eventually resulting in a transition to a very different regime, marked by strong e-e interactions and suppressed ST splittings.

The paper is organized as follows. 
A summary of symbols and abbreviations is presented in Table~\ref{tab:symbol_list}. 
In Sec.\ \ref{sec:dot}, we describe the two-electron quantum dot Hamiltonian, and the TB and FCI methods used to solve it. 
An overview of our main results is presented in Sec.\ \ref{subsec:ow}, with further details described in the remainder of the section.
In Sec.\ \ref{sec:conclusion}, we summarize our results and conclude. 
The Appendices provide technical details about the methods used.

\section{\label{sec:dot}Methods}
This section presents the system Hamiltonian and calculational methods used in this work.  Sec.~\ref{subsec:Hamiltonian} presents the Hamiltonian, Sec.~\ref{subsec:single_electron} describes the TB methods used to compute the single-electron eigenstates, and Sec.~\ref{subsec:two_electron} describes the FCI methods used to compute the two-electron eigenstates.

\subsection{\label{subsec:Hamiltonian}Hamiltonian} 
\begin{figure}
\includegraphics[width=3.4 in]{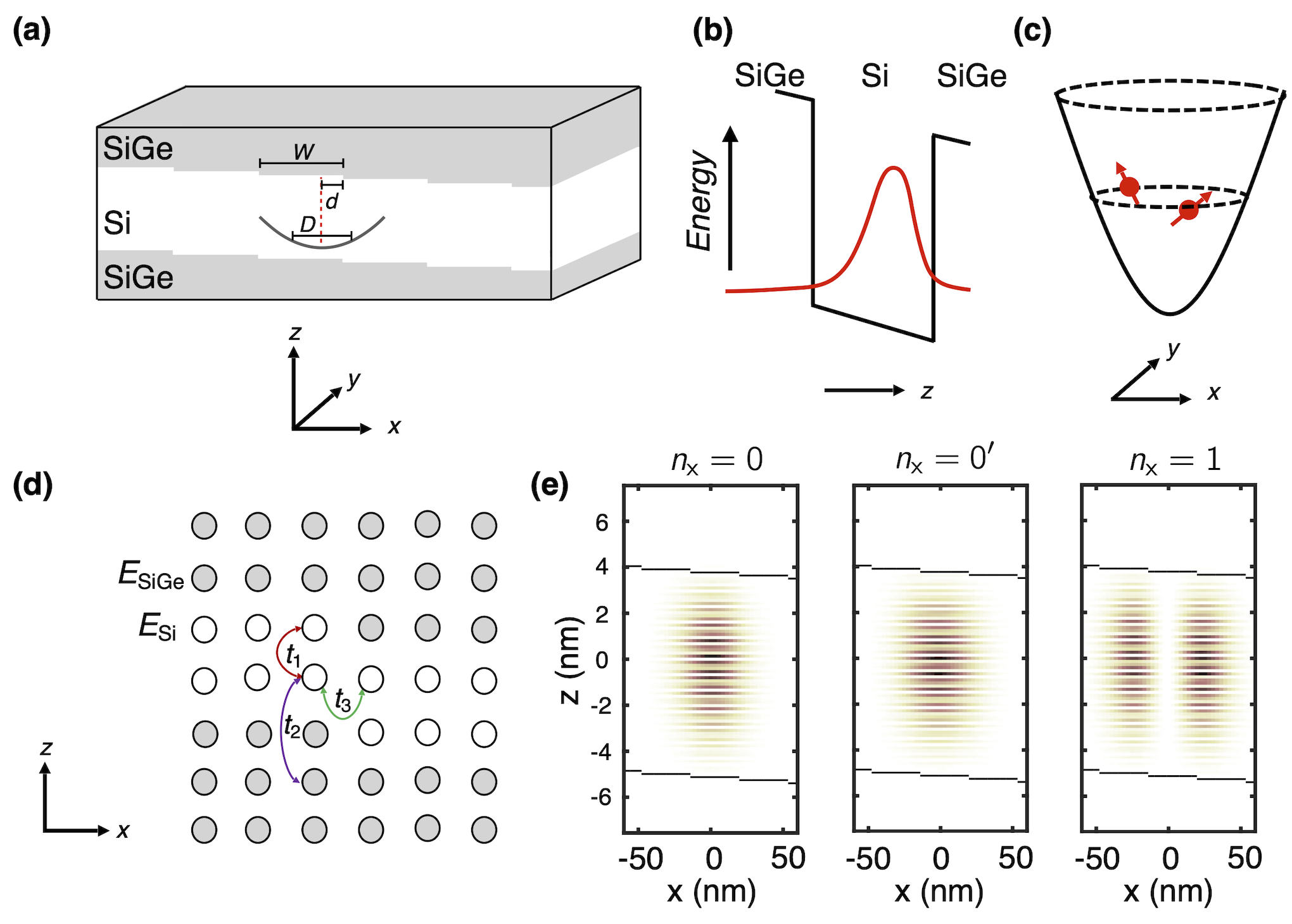}
\caption{Simulation and TB geometries used for single-electron calculations. 
(a) Schematic of the Si/SiGe heterostructure in the vicinity of the quantum well. 
The Si quantum well is depicted as white, while the SiGe barriers are depicted as gray.
Interfacial disorder is represented as single-atom steps of width $W$.
A quantum dot of diameter $D$ is formed within the quantum well, with separation $d$ between the center of the dot and the nearest step edge.
(b) Band offsets at the heterostructure interface, combined with an applied electric field perpendicular to the interface, confine the electrons vertically in the quantum well. The red curve represents the $z$-envelope of an electron wavefunction. 
Fast valley oscillations of the wavefunction are not shown. 
(c)  Electrostatic confinement in the $x$-$y$ plane is provided by top gates (not shown).
We approximate the 2D potential as harmonic. 
(d) TB model and hopping parameters $t_1$, $t_2$, and $t_3$ in the $x$-$z$ plane. 
On-site parameters for SiGe (gray) and Si (white) allow for the inclusion of atomic steps at the interface.
Note that the lattice shown here is only schematic; the number of TB sites used in the calculations is actually much higher ($\sim 10^4$). 
(e) Typical TB density results for the ground state ($n_\mathrm{x}=0$), the first valley-excited state ($n_\mathrm{x}=0'$), and the first orbitally excited state ($n_\mathrm{x}=1$). 
Although atomic steps at the interface cause VOC, it is usually possible to distinguish valley-like and orbital-like excitations.
Note that confinement along $\hat y$ is treated independently here, due to the separation of variables, and that excited $z$ subbands do not enter our calculations as their energies are very high.
Also note that the wavefunctions are plotted with different scales in the $\hat x$ and $\hat z$ directions.
}
\label{fig:fig1L}
\end{figure}

We consider a quantum dot containing two electrons, with Hamiltonian 
\begin{equation}
H^{\mathrm{2e}}=H^{\mathrm{1e}}(\bm{r}_1)+H^{\mathrm{1e}}(\bm{r}_2)+H_\text{int}(\bm{r}_1,\bm{r}_2),
\label{H2e}
\end{equation} 
where $H^{\mathrm{1e}}$ are the single-electron Hamiltonians, $H_\text{int}=\frac{e^2}{4 \pi \epsilon_0 \epsilon_r } \frac{1}{\abs{\bm{r}_1-\bm{r}_2}}$ is the Coulomb interaction, $e$ is the electron charge (with $e>0$), $\epsilon_0$ is the vacuum permittivity, and $\epsilon_r=11.4$ is the dielectric constant of low-temperature Si~\cite{Faulkner:1969p713}. 
Here, $\hat z$ is defined as the crystallographic [001] axis, although the quantum-well interface may not be perfectly aligned with $\hat z$, as depicted in Fig.~\ref{fig:fig1L}(a). 

The single-electron Hamiltonians are defined as
\begin{equation}
H^{\mathrm{1e}}=H_{\mathrm{K}}+H_{\mathrm{E}}+H_{\mathrm{QW}},
\end{equation}
where the kinetic energy $H_\text{K}$, the lateral confinement potential $H_\text{E}$, and the vertical confinement potential $H_\text{QW}$ are described below.

To account for VOC and atomic-scale disorder, we adopt an atomistic Hamiltonian in the $x$-$z$ plane.
We specifically consider the empirical two-band TB model of Refs.~\cite{Boykin:2004p115,Abadillo-Uriel:2018p165438}, as illustrated in Fig.~\ref{fig:fig1L}(d), which is known to give valley splitting results comparable to more realistic models~\cite{Boykin:2004p115,Boykin:2004p165325,Friesen:2007p115318}.
To improve the computational efficiency of our calculations, we adopt a continuous Hamiltonian in the third dimension (along $\hat y$). 
The only constraint imposed by this choice is that the interfacial disorder profile (e.g., the atomic steps) must be uniform along $\hat y$, as shown in Fig.~\ref{fig:fig1L}(a); this constraint does not affect any of our conclusions. 
By choosing additive confinement potentials of the form, $H^{1e}=H^{1e}_{\mathrm{x,z}}+H^{1e}_{\mathrm{y}}$, the single-electron calculations become separable.
The kinetic energy is thus given by
\begin{multline}
H_{\mathrm{K}}=\frac{-\hbar^2}{2m_t} \frac{\partial^2}{\partial y^2}  +\sum_{i,j} (t_1 \ket{i,j+1} \bra{i,j} \\ + t_2 \ket{i,j+2} \bra{i,j} + t_3 \ket{i+1,j}\bra{i,j} + \text{h.c.}) ,
\end{multline}
where the integer indices $i$ and $j$ refer to TB sites along the $\hat x$ and $\hat z$ axes, respectively.
The $\hat z$ hopping parameters, $t_1=0.68$~eV and $t_2=0.61$~eV, are chosen to give the correct longitudinal (out-of-plane) effective mass for Si, $m_l=0.916 \ m_0$, and the correct positions of the valley minima in the Brillouin zone, $k_0=\pm 0.82 (2\pi/a)$, where $a=5.43$ $\text{\normalfont\AA}$ is the Si cubic lattice constant, and the grid spacing is taken to be $\Delta z=a/4$~\cite{Boykin:2004p165325}.
The $\hat x$ hopping parameter, $t_3=-0.026$~eV, is chosen to yield the correct transversal (in-plane) effective mass for silicon, $m_t=0.191 \ m_0$.
We choose a grid spacing of $\Delta x=2.79$~nm for most of our calculations~\footnote{The exception is the $\hbar \omega \geq 10$~meV region of Fig.~\ref{fig:fig3L}(c). In this case, the grid spacing, $\Delta x$, is chosen to be smaller (0.47~nm, yielding $t_3=-0.93$~eV), since the confinement strength is stronger for this calculation.}. 
We note that grid spacings along $\hat x$ can be much larger than along $\hat z$, as there are no fast valley oscillations along $\hat x$.

The quantum well confinement potential, as illustrated in Fig.~\ref{fig:fig1L}(b) is defined as
\begin{multline}
H_{\mathrm{QW}}=\sum_{i,j} \Big[E_0+V_{\mathrm{QW}} \Theta_{i,j} \\ - e(iF_x \Delta x+jF_z  \Delta z)\Big] \ket{i,j} \bra{i,j}.
\end{multline}
Here, $E_0$ is a uniform energy shift, $V_{\mathrm{QW}}=150$ meV is a typical band offset between Si and SiGe, and $\Theta_{i,j}$ is a step function that takes the value $1$ on a SiGe site and $0$ on a Si site.
$\bm{F}=(F_x,0,F_z)$ is a uniform electric field produced by a top-gate electrode, oriented perpendicularly to the average quantum-well interface.
In this work, we considered single-atom steps that are either equally spaced, or isolated. 
The first case results in an average interface tilt that determines the orientation of $\bm{F}$. 
We refer to such disorder as a ``tilted interface." 
For the case of an isolated step, we simply take $\bm{F}$ to be parallel to $\hat z$.
In both cases, the choice of step disorder applies to all the interfaces in the device.

Finally, the electrostatic confinement potential is assumed to be harmonic along both $\hat x$ and $\hat y$:
\begin{multline}
H_{\mathrm{E}}=\frac{1}{2}m_t \omega_x^2 \sum_{i,j} (i\Delta x)^2 \ket{i,j} \bra{i,j} + \frac{1}{2}m_t \omega_y^2 y^2,
\end{multline}
as illustrated in Fig.~\ref{fig:fig1L}(c).

\subsection{\label{subsec:single_electron}Single-electron wavefunctions} 
The first step in our solution procedure is to compute the single-electron energy eigenstates.
We therefore solve $H^{\mathrm{1e}} \phi_i=\varepsilon_i \phi_i$ by leveraging the separability of the Hamiltonian, $H^{1e}=H^{1e}_{\mathrm{x,z}}+H^{\mathrm{1e}}_{\mathrm{y}}$. 
The TB Hamiltonian $H^{1e}_{\mathrm{x,z}}$ is numerically diagonalized to obtain the wavefunctions $\zeta_{n_{x,z}}(x,z)$, where the labels $n_{x,z}$ denote the valley and orbital modes. 
We choose the spatial domain in the $x$-$z$ plane to be large enough that the wavefunctions have already vanished before reaching the boundaries. Some typical wavefunction solutions for $|\zeta_{n_{x,z}}(x,z)|^2$ are shown in Fig.~\ref{fig:fig1L}(e) for a tilted interface. 
Note that confinement along $\hat z$ is strong enough that only the lowest subband is relevant for our calculations. 
As $H^{\mathrm{1e}}_{\mathrm{y}}$ is taken to describe a harmonic oscillator in this work, the corresponding wavefunctions $\eta_{n_y}(y)$ may be obtained analytically, as described in Appendix~\ref{sec:app1b}.
The full single-electron wavefunctions $\phi_{n_{x,z},n_y}(x,y,z)=\zeta_{n_{x,z}}(x,z)\eta_{n_y}(y)$ are then employed as basis states in the FCI calculations, as described below.

\subsection{\label{subsec:two_electron}Two-electron wavefunctions} 
The single-electron wavefunctions, computed above, represent exact solutions in the absence of Coulomb interactions, $H_\text{int}(\bm{r}_1,\bm{r}_2)$.
We treat the strong e-e interactions of interest here nonpertubatively by using the well-known FCI procedure to solve for the two-electron wavefunctions~\cite{Szabo:1996p91861}.
This exact diagonalization technique is frequently used in quantum chemistry applications and has also been used to study quantum dot systems~\cite{Bryant:1987p1140,Hu:2000p62301,Reimann:2002p1283,Friesen:2003p121301,Korkusinski:2007p115301,Singha:2010p246802,Nielsen:2010p75319,Barnes:2011p235309,Nielsen:2012p114304,Wang:2010p235326,Liu:2012p45311,Bakker:2015p155425,Pan:2020p34005,Buonacorsi:2020preprint}. 

\begin{figure*}
\includegraphics[width=7in]{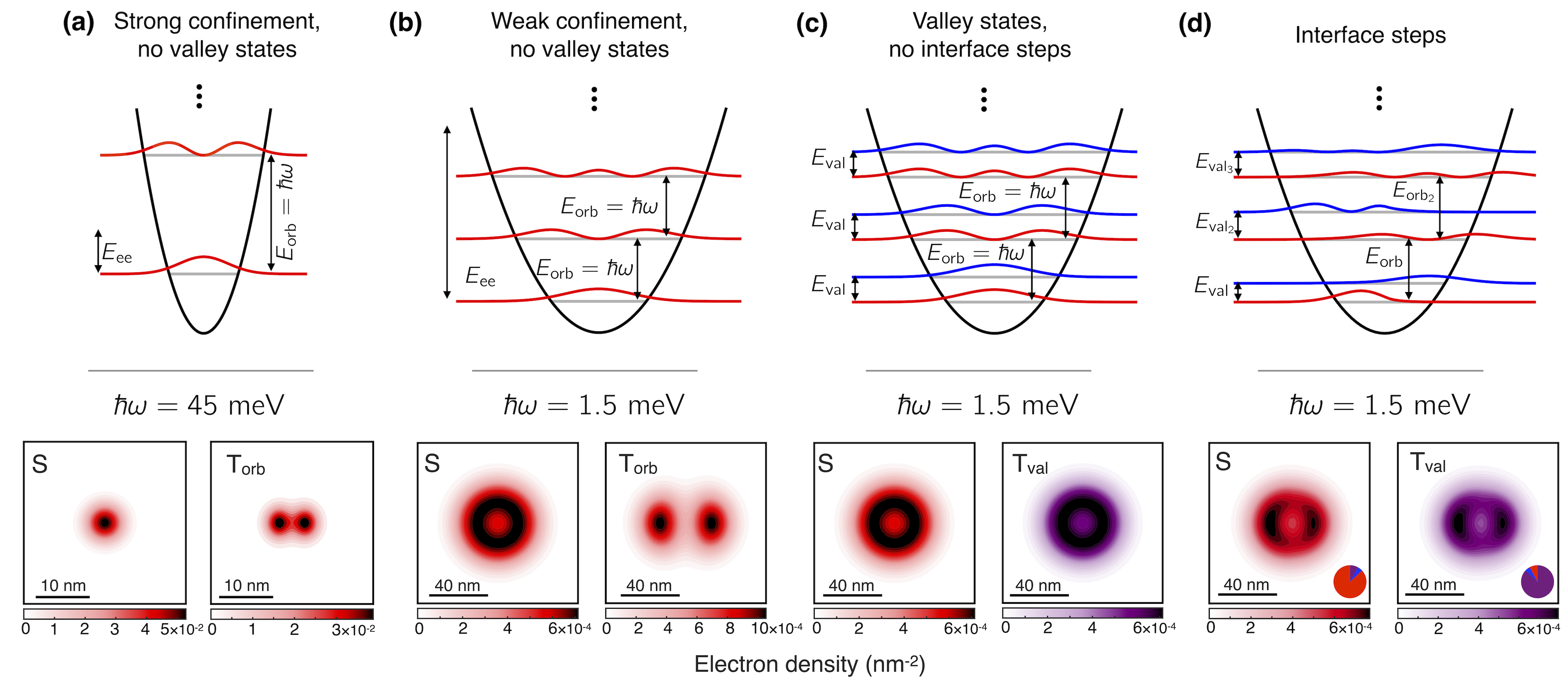}
\caption{Summary of main results. 
Top row shows schematic single-electron wavefunctions.
New physical ingredients, indicated by labels, are introduced into the calculations, one by one.
The relations between different energy scales are also indicated:
(a) $E_\text{ee} < E_\text{orb}$ (no valleys); (b) $E_\text{orb} \ll E_\text{ee}$ (no valleys); (c) $E_\text{val} < E_\text{orb} \ll E_\text{ee}$ (no interface steps); (d) $E_\text{val} < E_\text{orb} \ll E_\text{ee}$ (dot centered at an interface step).
Color coding: orbital excitations (red), valley excitations (blue).
Second row shows two-electron densities, obtained from FCI calculations.
Confinement energies are specified ($\hbar\omega$), and the vertical electric field is (a) 40~MV/m, (b-d) 1.5~MV/m.
Some distinctive features include
(a) Wavefunctions dominated by low-energy configurations; (b) Wigner molecules shaped like (S) a doughnut, or (T$_\text{orb}$) a dumbbell; (c) S and T$_\text{val}$ have identical shapes; (d) A single interface step breaks the circular confinement symmetry.
Color coding: same-valley eigenstates (red or blue), opposite-valley eigenstates (red + blue = purple).
In (d), the broken valley symmetry induced by interfacial disorder causes orbital and valley states to hybridize, with the weights of same-valley and opposite-valley configurations indicated in the pie charts.
}
\label{fig:fig2L}
\end{figure*}

In the FCI approach, we use a complete basis set of single-electron wavefunctions $\{\phi_i\}$ to construct a complete basis set of two-electron wavefunctions $\{\psi_{\alpha}\}$. 
The basic procedure has two steps. 
First, we combine the spatial wavefunctions with spin wavefunctions to form spin orbitals. 
Here, we use the shorthand notation $\chi_{2i}=\phi_i \otimes \uparrow$ and $\chi_{2i+1}=\phi_i \otimes \downarrow$. 
Next, we combine two of these spin orbitals and ensure that they have proper fermionic symmetry by computing their Slater determinants, $\psi_{\alpha}=\ket{\chi_m \chi_n}$, using the notation of Ref.~\cite{Szabo:1996p91861} to denote Slater determinants. 
The resulting states are referred to as configurations.
The set of all possible configurations, obtained from all possible spin-orbital combinations, forms a complete basis set for expanding the two-electron Hamiltonian.
A general eigenstate $\Psi$ of $H^{2e}$ can therefore be expressed as  
\begin{equation}
\Psi = \sum\limits_{\alpha=0}^{\infty}c_{\alpha} \psi_{\alpha}.
\label{SlaterExp}
\end{equation}
Slater determinants are not necessarily spin eigenstates; however it is straightforward to combine configurations to obtain such spin-adapted configurations, $\psi_\alpha^\text{(ST)}$~\cite{Szabo:1996p91861}. 
Hence, we can also write
\begin{equation}
\Psi = \sum\limits_{\alpha=0}^{\infty}c_{\alpha}^{(\mathrm{ST})} \psi_{\alpha}^{(\mathrm{ST})}.
\label{STExp}
\end{equation}
In practice, it is not possible to work with an infinite basis set.
We therefore truncate the single-electron basis states to a set of size $K$, which is large enough that the results are not significantly affected when $K$ is increased. 
Typically, we find that $K=42$ is sufficient for our simulations, corresponding to six single-electron orbital shells.
The full calculation therefore involves $2K$ spin orbitals and ${2K\choose{2}}=3486$ configurations. 

Calculating the $H^{2e}$ matrix elements involves solving two types of integrals: $\bra{\psi_{\alpha}} H^{1e}(\bm{r}_i) \ket{\psi_{\beta}}$ and $\bra{\psi_{\alpha}} H_{\mathrm{int}} \ket{\psi_{\beta}}$. 
The first type can easily be reduced to single-electron integrals whose values are known from the single-electron calculations. 
The remaining Coulomb integrals are, by far, the most computationally expensive part of the simulation, since the number of unique integrals scales as $K^4$. 
It is therefore desirable to identify an efficient computational method for performing these integrals.
Here, we adopt a scheme~\cite{Molnar:2002p7795} that allows us to exploit fast Fourier transform (FFT) algorithms, which are numerically efficient. 
Additionally, having incorporated separability into our single-electron Hamiltonians, we can exploit the harmonic oscillator wavefunctions $\eta_{n_y}(y)$ to compute the $y$ integrals analytically. 
This allows us to replace the 3D Coulombic interaction $1/r$ with an effective 2D interaction $\tilde{f}$ in the $k_x$-$k_z$ Fourier plane.
The resulting electrostatic potential for one of the electrons is then given by
\begin{multline}
V(x,z)=\frac{e}{4 \pi \epsilon_0 \epsilon_r}\int dk_xdk_z e^{2 \pi i(k_xx+k_zz)} \\ \times \tilde{f}(k_{x},k_{z}) \tilde{\rho}^{(1)}(k_x,k_z),
\end{multline} 
where $\tilde{\rho}^{(1)}(k_x,k_z)$ is the Fourier transform of $\rho^{(1)}(x,z)\equiv\zeta_i^*(x,z)\zeta_j(x,z)$, as obtained from the discrete TB calculation.
Here, $V(x,z)$ and $\tilde{\rho}^{(1)}(k_x,k_z)$ are both computed using an FFT routine. 
The full interaction integral then has the form $e\int dx dz V(x,z) \rho^{(2)}(x,z)$, which we perform as a discrete summation in real space. 
Our FCI computational strategy is summarized in more detail in Appendix~\ref{sec:app1}, while the details of the FFT-based integration method are described in Appendix~\ref{sec:app1b}.  

\section{\label{sec:ST}Results} 
In this section, we present our results for two-electron wavefunctions and ST splittings.  
We first provide an overview of the physics of two-electron wavefunctions in a Si quantum dot, followed by a more focused discussion of behavior observed as a function of confinement, for both circular and elliptical dots, and for dots with different disorder profiles. 
Geometries without interfacial disorder (i.e., without VOC) are easier to understand, and they provide a baseline for interpreting the more realistic, yet complicated, behaviors observed when VOC is present. 

\subsection{\label{subsec:ow}Overview of the physics}

Here, we provide an overview of the physics of two-electron dots, including the effects of e-e interactions and interfacial disorder.
We begin with the simplest case of weak e-e interactions, with no valleys and no disorder.
(Note that, since the relative strength of e-e interactions is governed by $E_{\mathrm{ee}}/E_{\mathrm{orb}}$, 
the weakly interacting regime may also be viewed as the strongly confined regime; we use the two terminologies interchangeably.)
We then add each of these ingredients to our calculations, one at a time, to achieve a realistic description of the two-electron system. 
A schematic summary is presented in Fig.~\ref{fig:fig2L}.

A system with weak e-e interactions, no valleys, and no disorder is easy to describe.  
Such situations can arise when the valley and orbital splittings are both much larger than other experimental energy scales.
In the noninteracting limit, the two-electron wavefunctions are well-known from quantum mechanics textbooks, and are equivalent to low-energy FCI configurations:
the ground state is a spin singlet with spatial configuration given by $\phi_0(\mathbf{r}_1)\phi_0(\mathbf{r}_2)$, and
the low-energy spin triplet has the spatial configuration $[\phi_0(\mathbf{r}_1)\phi_1(\mathbf{r}_2)-\phi_1(\mathbf{r}_1)\phi_0(\mathbf{r}_2)]/\sqrt{2}$.
For the noninteracting case, the ST energy splitting is therefore given by $E_\text{ST}=E_\text{orb}=E_1-E_0$.
In Fig.~\ref{fig:fig2L}(a), we show (schematically) the low-energy single-electron basis states, $\phi_0,\phi_1,\dots$,
and the corresponding probability densities of the two-electron singlet and triplet states.
For the triplet, we note that $\phi_1$ is doubly degenerate (e.g., $p_x$ vs.\ $p_y$), leading to degenerate triplets, of which we have only shown one.

We now include strong e-e interactions,  with $E_\text{ee}\gg E_\text{orb}$, but still no valleys.
Such a situation can arise when the dot is large.
In this regime, the energy cost for adding new configurations to the two-electron wavefunctions is low compared with the potential benefits of using these configurations to reduce $E_\text{ee}$.
This leads to the formation of Wigner molecules, as shown in Fig.~\ref{fig:fig2L}(b) for the low-energy singlet and triplet.
For the singlet case, we note that the circular symmetry of the confinement potential is also conferred upon the two-electron wavefunction, yielding a Wigner molecule shaped like a doughnut.
In general, the configurations that make up singlet wavefunctions have very different probability distributions than those of triplet wavefunctions, beginning already with the lowest-energy configurations, as described above.
This is apparent in the two-electron wavefunctions shown in Fig.~\ref{fig:fig2L}(a). 
However in the strongly interacting limit, when a very large number of configurations contribute to each solution, the singlet and triplet wavefunctions grow more similar in appearance and in energy.
This is because minimizing $E_\text{ee}$ is the driving force for both the singlet and triplet calculations, and when many degrees of freedom are available for doing this, the resulting charge distributions take the same shape.
Consequently, $E_\text{ST}$ is strongly suppressed in the weakly confined regime, and ultimately approaches zero.

Next, we add valleys, but not interfacial disorder. 
For definiteness, we assume that $E_\text{val}<E_\text{orb}$, which is typical for most experiments.
As seen in Fig.~\ref{fig:fig2L}(c), each orbital state now splits into a pair of states (red and blue) with the same wavefunction envelope.
The envelopes are modulated by fast valley oscillations (not shown), which are 90 degrees out of phase, yielding appropriately orthogonalized wavefunctions~\cite{Friesen:2007p115318}. 

Due to the lack of interfacial disorder, the valley degree of freedom remains a good quantum number.
Thus for strongly interacting electrons, the two-electron wavefunctions are formed of many configurations that must all have the same valley character: either two electrons in the same valley state, or two electrons with opposite valley states.
In particular, the ground-state singlet is formed of same-valley configurations.
However, there are now two types of low-energy triplets: The first is formed of same-valley configurations, which we refer to as an orbital triplet, $\mathrm{T_\text{orb}}$; the second is formed of opposite-valley configurations, which we refer to as a valley triplet, $\mathrm{T_\text{val}}$.
In the following sections, we analyze the ST splittings obtained for these two triplet states, defined as $E_{\text{ST}_\text{orb}}$ and $E_{\text{ST}_\text{val}}$.

Generally, the S and T$_\text{orb}$ states have the same shape and behavior as the S and T states described above, for the case where no valleys were considered.
However, T$_\text{val}$ behaves differently.
To understand this, we first note that ``valley" is a good quantum number for single-electron wavefunctions with no disorder.
This symmetry also extends to two-electron solutions, except that the Coulomb interaction term, $H_\text{int}$, breaks the symmetry very slightly.
In the following section, and in Appendix~\ref{sec:app2}, we clarify the weak effects of this symmetry breaking.
However, for the present purposes, we simply note that there is an approximate symmetry ensuring that $\text{T}_\text{val}$ has the same configuration expansion as S, except with each spatially symmetric configuration replaced by a spatially antisymmetric configuration, with one of the lower valley states replaced by an excited valley state.
This (approximate) symmetry is apparent when we compare the S and T$_\text{val}$ wavefunctions in Fig.~\ref{fig:fig2L}(c).
Here, the S wavefunction is formed exclusively of same-valley configurations (its probability density is colored red) and the T$_\text{val}$ probability density is formed of opposite-valley configurations (colored red + blue = purple).
However, the shapes of the probability distributions appear identical.
In addition to their nearly identical shapes, the wavefunction energies are nearly identical, with $E_{\text{ST}_\text{val}}\approx E_\text{val}$~\footnote{This result is consistent with the perturbative calculations of Ref.\ \cite{Jiang:2013p85311}}.
This intriguing and important result will be explored in detail below.

Finally, we consider the most physically relevant scenario, which includes disorder at the quantum well interface, which induces VOC and breaks the valley symmetry. 
As a result of VOC, the orbital and valley characters of the single-electron basis states hybridize, yielding pairs of valley states with dissimilar envelopes. This behavior is depicted at the top of Fig.~\ref{fig:fig2L}(d).
(Note that the single-electron envelopes in the figure are merely schematics, which we have exaggerated for visual clarity.)
The phases of the fast valley oscillations (not shown) are also affected by VOC, and vary from orbital to orbital.
Despite the VOC, we note that it is usually possible to identify valley-like or orbital-like excitations in our TB wavefunctions, particularly when $E_\text{val}$ is well separated from $E_\text{orb}$.
In the remainder of this paper, we therefore refer to them simply as valley or orbital states, when it makes sense to do so. 

\begin{figure*}
\includegraphics[width=7 in]{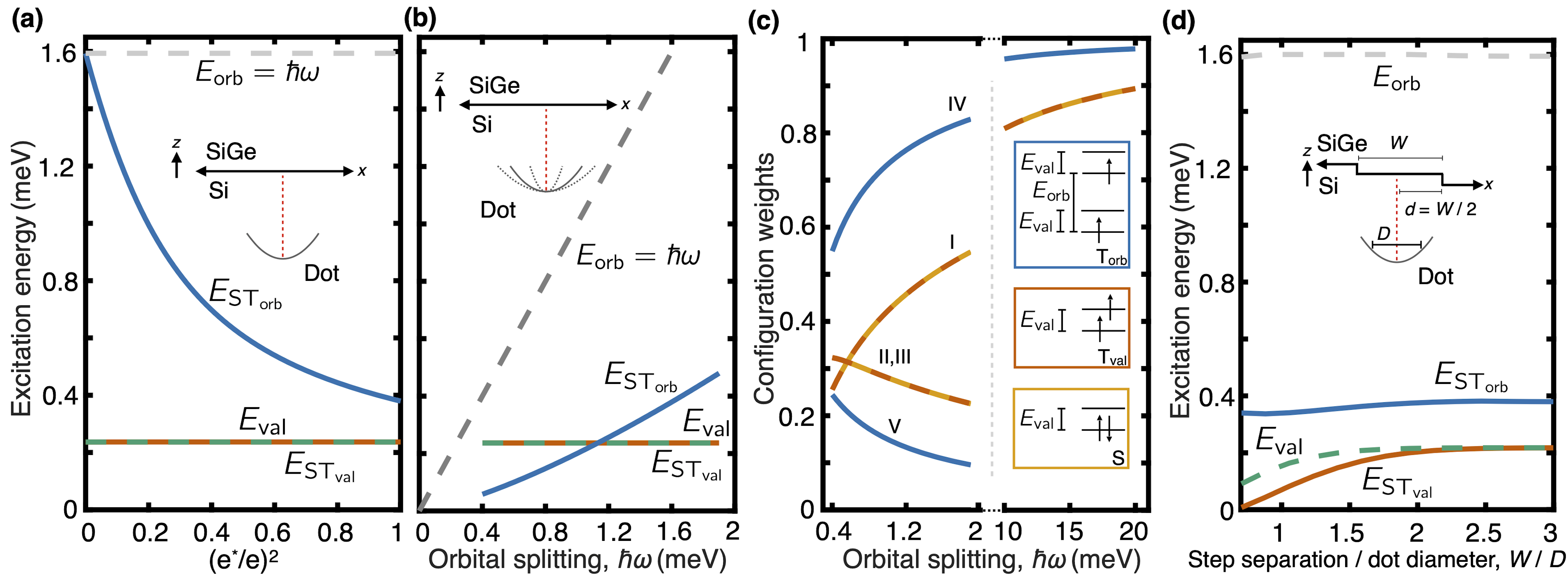}
\caption{
Effects of e-e interactions and valley-orbit coupling (VOC) in silicon quantum dots.
(a), (b) Comparison of the single-electron valley splitting (cyan-dashed) and orbital splitting (gray-dashed) to the two-electron orbital (blue) and valley (red) singlet-triplet splittings. 
Here, there is no step disorder at the quantum well interface, and the e-e interactions are controlled by (a) the artificially modified Coulomb interaction strength $(e^*/e)^2$, with $E_\text{orb}=\hbar\omega=1.6$~meV, or (b) the orbital confinement strength $\hbar\omega$, with $e^*=e$.
In both cases, $E_{\text{ST}_\text{orb}}$ is suppressed well below its noninteracting value $E_\text{orb}$, as consistent with strongly interacting electrons, while $E_{\mathrm{ST_{val}}}\approx E_\text{val}$ is essentially unaffected by e-e interactions.
In (b), the lowest triplet state switches from T$_\text{orb}$ (low $\hbar\omega$), whose energy depends strongly on the confinement, to T$_\text{val}$ (high $\hbar\omega$), whose energy is nearly the same as S.
(c) Weights of the low-energy configurations, I-V in Table~\ref{tab:table1}, for the same S, T$_\text{orb}$, and T$_\text{val}$ solutions shown in (b).
The weights shown for S and T$_\text{val}$ are nearly identical, as consistent with $E_{\mathrm{ST_{val}}}\approx\text{constant}$.
For very strong confinement, the wavefunctions converge slowly to their lowest configurations (depicted in the insets, along with the color coding),  indicating that the electrons are strongly interacting throughout the normal operating regime.
(d) Excitation energies for a dot with VOC, induced by a tilted interface.
Here, $E_\text{orb}=\hbar\omega=1.6$~meV corresponds to a dot of diameter $D=2\sqrt{\hbar/m^* \omega}=31.7$~nm, centered halfway between two steps ($d=W/2$).
$E_{\mathrm{ST_{val}}}$ is largest when the step separations are very large.
For all panels, the quantum well width is 10~nm, the barrier height between Si and SiGe is 150~meV, the dot is circularly symmetric, and the vertical electric field is 1.5~MV/m, for the experimentally relevant range $\hbar \omega \leq$~2 meV.
In the very highly confined regime of (c) ($\hbar \omega \geq$~10 meV), we use a high electric field, 40~MV/m, to avoid filling higher subbands.}
\label{fig:fig3L}
\end{figure*}

The broken valley symmetry extends to the two-electron states shown in Fig.~\ref{fig:fig2L}(d), since the configurations are no longer strictly same-valley or opposite-valley.
The presence of interfacial steps also breaks the circular symmetry of the confinement potential, so the probability densities take a shape intermediate between a doughnut and a dumbbell.
The color of the probability density indicates that the S wavefunction is mainly formed of same-valley configurations (red with a blue tinge).
Similarly, the T$_\text{val}$ probability density is formed mainly of opposite-valley configurations (purple with a red tinge).
The actual same-valley (red) vs.\ opposite-valley (purple) configuration weights are specified in the pie charts shown as insets; in all the other panels, the colors are pure-red or pure-purple.

Finally, we note that, since the approximate valley symmetry protecting $E_{\text{ST}_\text{val}}$ is broken by the interfacial disorder, it is no surprise that $E_{\text{ST}_\text{orb}}$ and $E_{\text{ST}_\text{val}}$ are now both affected by e-e interactions. The ST splittings for dots with VOC will be discussed in detail in later sections.

\subsection{Characterizing e-e interactions}\label{sec:eeinteractions}
This section provides an in-depth analysis of the effects of e-e interactions.
To begin, we return to the case of no interfacial disorder, and therefore, no VOC.

We first recall that the ratio $E_\text{ee}/E_\text{orb}$ characterizes the strength of e-e interactions.
However, the experimental tuning knob for the confinement energy, $\hbar\omega$, affects both $E_\text{orb}$ and $E_\text{ee}$.
We can disentangle these parameters theoretically by holding $\hbar\omega$ fixed while tuning the effective charge used in the calculations, $e^*$.
In Fig.~\ref{fig:fig3L}(a), we plot several excitation energies as a function of the effective Coulomb interaction strength, $(e^*/e)^2$, for typical quantum dot parameters, in the absence of interfacial disorder.
When $e^*=0$, we observe results consistent with the noninteracting limit, described above, for which $E_{\mathrm{ST_{orb}}}=E_\text{orb}$.
When $e^*>0$, $E_{\mathrm{ST_{orb}}}$ is quickly suppressed, which can be understood by considering the lowest-energy configurations for S, T$_\text{val}$, and T$_\text{orb}$, as depicted in the insets of Fig.~\ref{fig:fig3L}(c).
Since both electrons occupy the same orbital for S, and since this is the smallest orbital, the Coulomb energy is greater for S than it is for T$_\text{orb}$.
Therefore, as $(e^*/e)^2$ rises, the energy splitting $E_{\mathrm{ST_{orb}}}$ decreases.
The trend continues as the interaction strength grows, and many configurations begin to contribute to the wavefunctions.
The fact that $E_{\mathrm{ST_{orb}}}$ reaches a value for which $E_{\mathrm{ST_{orb}}}\ll E_\text{orb}$ at full interaction strength, $e^* = e$, indicates that the electrons are strongly interacting.

Turning to T$_\text{val}$, we see that $E_{\mathrm{ST_{val}}}=E_\text{val}$ in the noninteracting limit, $e^*=0$, as expected.
However, in stark contrast with the behavior of $E_{\mathrm{ST_{orb}}}$, we find that $E_{\mathrm{ST_{val}}}\approx E_\text{val}$ over the whole range of coupling strengths.
As discussed in the previous section, this remarkable behavior can be attributed to the valley symmetry present in the absence of VOC, and persists (approximately), even when the interactions are strong.

The nature of the valley symmetry (and symmetry breaking) can be understood by taking a closer look at the Coulomb matrix elements giving rise to the weak symmetry breaking.
Full details of this discussion are presented in Appendix~\ref{sec:app2}.
Here, we summarize those findings by considering the single-electron spatial wavefunctions $\phi_\alpha(\mathbf{r})$ and ${\phi_\alpha}'(\mathbf{r})$, where ${\phi_\alpha}'(\mathbf{r})$ refers to the valley-excited version of the $\phi_\alpha(\mathbf{r})$ orbital.
To a good approximation, $\phi_\alpha(\mathbf{r})$ and ${\phi_\alpha}'(\mathbf{r})$ have the same envelopes, but their fast valley oscillations differ in phase by 90 degrees~\cite{Friesen:2007p115318}.
Because of these fast oscillations, we expect that
\begin{equation}
\int d\mathbf{r}_1 \phi_\alpha (\mathbf{r}_1) {\phi^*_\beta}' (\mathbf{r}_1) V(\mathbf{r}_1,\mathbf{r}_2)\approx 0 ,
\end{equation}
in the regimes studied here, for all $\alpha,\beta$, whenever $V(\mathbf{r}_1,\mathbf{r}_2)$ is a slowly varying function. 
However, $V(\mathbf{r}_1,\mathbf{r}_2)$ is the Coulomb interaction, and is therefore singular when $\mathbf{r}_1=\mathbf{r}_2$, yielding matrix elements that are nonvanishing.
Consequently, the valley symmetry is lifted by Coulomb interactions.
However, the resulting matrix elements are orders of magnitude smaller than matrix elements for which both single-electron wavefunctions are of the same-valley type.
A second type of integral plays a role in ensuring that $E_{\mathrm{ST_{val}}}\approx E_\text{val}$:
\begin{multline}
\int d\mathbf{r}_1 \phi_\alpha (\mathbf{r}_1) {\phi^*_\beta} (\mathbf{r}_1) V(\mathbf{r}_1,\mathbf{r}_2)
\\ \approx 
\int d\mathbf{r}_1 {{\phi_\alpha}' (\mathbf{r}_1)} {\phi^*_\beta}' (\mathbf{r}_1) V(\mathbf{r}_1,\mathbf{r}_2) .
\end{multline}
In this case, the approximate equality occurs because $\phi_\alpha(\mathbf{r})$ and ${\phi_\alpha}'(\mathbf{r})$ have nearly identical envelopes~\cite{Hada:2003p155322}.

\begin{table*}
\caption{\label{tab:table1}
Lowest-energy configurations comprising S, $\mathrm{T_{val}}$, and $\mathrm{T_{orb}}$.
The notation $(n_x, n_y ;n_{\bar{x}}  , n_{\bar{y}} )$ denotes properly normalized and symmetrized (or anti-symmetrized) spatial wavefunctions, formed as products of the single-electron states $\phi_{n_x,n_y}(\mathbf{r}_1)$ and $\phi_{n_{\bar{x}},n_{\bar{y}}}(\mathbf{r}_2)$.
Here, $n_x$ ($n_y$) denotes the orbital quantum numbers for harmonic confinement along $\hat x$ ($\hat y$), and the primed notation indicates an excited valley state. 
Configurations I (S and T$_\text{val}$) and IV (T$_\text{orb}$) are depicted in the insets of Fig.\ \ref{fig:fig3L}(c).\vspace{.1in}} 
\begin{tabular}{c c c c c c c c c c c c}
\toprule
 &&& S &&&& T$_\text{val}$ &&&& T$_\text{orb}$ \\
 \colrule
I &&& $(0,0;0,0)$ &&&& $(0,0;0',0)$  &&&& - \\
II &&& $((1,0;1,0)+(0,1;0,1))/ \sqrt{2}$ &&&& $((1,0;1',0)-(0,1;0',1))/ \sqrt{2}$  &&&& - \\
III &&& $((0,0;2,0)+(0,0;0,2))/ \sqrt{2}$ &&&& $((0,0;2',0)+(0,0;0',2)-(0',0;2,0)-(0',0;0,2))/2$  &&&& - \\
IV &&& - &&&& - &&&& $(0,0;1,0)$ \\
V &&& - &&&& - &&&& $(1,0;2,0)$ \\
\botrule
\end{tabular}
\end{table*}

To further characterize e-e interactions, we plot excitation energies as a function of the orbital splitting in Fig.~\ref{fig:fig3L}(b), and we plot the corresponding configuration weights ($|c_\alpha^\text{ST}|^2$) in Fig.~\ref{fig:fig3L}(c), as defined in Eq.~(\ref{STExp}).
Again we observe a remarkable difference in behaviors between $E_{\mathrm{ST_{orb}}}$ and $E_{\mathrm{ST_{val}}}$.
Here, we also observe a crossing between the triplet states, which plays an important role in experiments.
Specifically, for large quantum dots (small $\hbar\omega$), the excitation energy is a strong function of $\hbar\omega$, while for small quantum dots (large $\hbar\omega$) it is not.
We define the minimum excitation energy as $E_\text{ST}=\text{min}(E_{\mathrm{ST_{val}}},E_{\mathrm{ST_{orb}}})$.

In Fig.~\ref{fig:fig3L}(c), we plot the weights of the dominant configurations contributing to S, T$_\text{orb}$, and T$_\text{val}$.
The spatial components of these configurations are listed in Table~II, with the lowest-energy configurations illustrated in the insets of the figure.
[In the large-$\hbar\omega$ region of Fig.~\ref{fig:fig3L}(c), a large vertical electric field was employed, to avoid populating higher subbands in the quantum well.]
Here, the valley symmetry causes the S and T$_\text{val}$ configuration weights to overlap, to a very good approximation.
There is additional degeneracy of levels II and III due to the circular symmetry of the confinement potential.
In the limit of very strong confinement (i.e., weak interactions) the two-electron wavefunctions are well approximated by their lowest-energy configurations, with the weights of all other configurations strongly reduced.
It is important to note, however, that a 90\% convergence of the weights is only achieved for $\hbar\omega>20$~meV.
For weaker confinement, we find that $E_{\mathrm{ST_{orb}}}\ll E_\text{orb}$.
This $\hbar\omega=20$~meV confinement level is much larger than for typical dots~\cite{Zajac:2016p054013,Chen:2020p044033,Dodson:2021}, including MOS dots~\cite{Yang:2012p15319}.
Moreover, the valley splittings in typical dots are also much lower than 20~meV.
Taken together, these results emphasize the relevance of e-e interactions and valley physics for all Si devices, and particularly for Si/SiGe quantum wells.

\subsection{Characterizing VOC}
\label{subsec:VOC}
In this section we explore the effects of single-atom steps at the quantum well interface, and the resulting VOC, for one and two-electron wavefunctions.

\begin{figure}
\includegraphics[width=3.4 in]{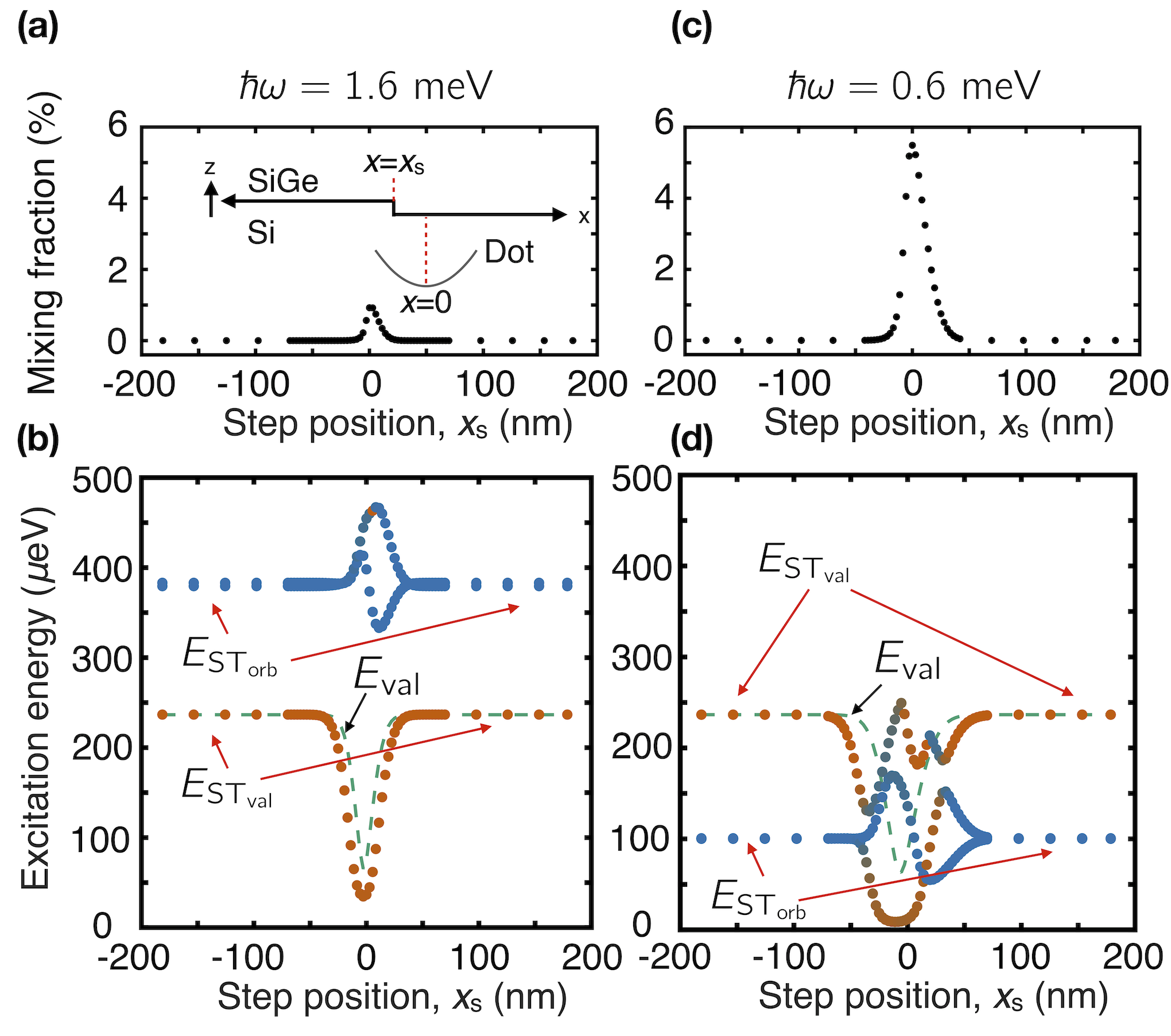}
\caption{
Effects of a single-atom step on valley splitting and VOC in one- and two-electron wavefunctions.
(a) and (c) correspond to a small dot. 
(b) and (d) correspond to a large dot.
In all cases, the quantum well is 10~nm thick and the vertical electric field is 1.5~MV/m.
(a),(b) Mixing fractions, as defined in Eq.~(\ref{Eq:mixing_fraction}), plotted as a function of the step position, as depicted in the inset.
A large mixing fraction indicates strong  VOC in single-electron dots.
(c),(d) Energy excitations as a function of the step position.
Dashed cyan lines show the valley splitting of the lowest orbital level.
Blue circles correspond to $E_{\mathrm{ST_{orb}}}$ and red circles correspond to $E_{\mathrm{ST_{val}}}$, where the colors reflect the total weight of same-valley configurations vs.\ opposite-valley configurations.
Here we plot two T$_\text{orb}$ energy levels, which are degenerate in the $|x_s|\gg 0$ regime.
The combination of VOC and e-e interactions suppresses $E_{\text{ST}_\text{val}}<E_\text{val}$, but has a mixed effect on $E_{\text{ST}_\text{orb}}$.
In (d), the triplet states anticross, causing the orbital vs.\ valley characters to switch.
}
\label{fig:fig4L}
\end{figure} 

\subsubsection{Isolated step}

Valley splitting is determined by the valley phase at the quantum well interface~\cite{Friesen:2007p115318}.
Since that phase differs by $\sim\pi$ on neighboring atomic layers, a single-atom step at the interface can cause significant interference effects that suppress the valley splitting and generate VOC~\cite{Friesen:2010p115324}.
The presence of VOC hybridizes valley states in different orbitals. 
Hence, the following mixing fraction, $M_0$, provides a direct measure of VOC in single-electron wavefunctions:
\begin{equation}
M_0=\sum_{m_{x,z}=1,1^{\prime},2,2^{\prime}, ...} \abs{\braket{\zeta_{0}}{\zeta_{m_{x,z}}^{\mathrm{(flat)}}}}^2 .
\label{Eq:mixing_fraction}
\end{equation}
Here, $\zeta_0(x,z)$ is the ground state TB wavefunction in the presence of an atomic step, $\zeta_{m_{x,z}}^\text{(flat)}(x,z)$ are TB wavefunctions computed in the absence of steps, and the sum is taken over all excited orbital states.
(Note that VOC only occurs in the $x$-$z$ sector of the wavefunctions, due to the separation of variables.)

We expect to observe larger VOC effects when a dot is closer to a step.
In Figs.~\ref{fig:fig4L}(a) and \ref{fig:fig4L}(c), we plot the mixing fraction as a function of the step location, $x_s$ (see inset for explanation), for two different dot sizes.
$M_0$ is clearly peaked for small $|x_s|$, and it is more strongly peaked for large dots, because the energy splittings are smaller, so many orbitals can contribute to VOC.
In Figs.~\ref{fig:fig4L}(b) and \ref{fig:fig4L}(d), we also see that $M_0$ is strongly correlated with the suppression of the valley splitting (dashed curves).

The effects of a step on two-electron excitation energies are also shown in Figs.~\ref{fig:fig4L}(b) and \ref{fig:fig4L}(d).
Here, red circles correspond to $E_{\text{ST}_\text{val}}$ and blue circles correspond to the two $E_{\text{ST}_\text{orb}}$ excitations, which are degenerate in the asymptotic regime, $|x_s|\gg 0$.
In the first case, the dot is small, so $E_\text{orb}>E_\text{val}$ and $E_{\text{ST}_\text{orb}}>E_{\text{ST}_\text{val}}$ in the asymptotic regime.
In the region near $x_s\approx 0$, $E_{\text{ST}_\text{val}}$ deviates from $E_\text{val}$ because VOC breaks the approximate valley symmetry that previously protected $E_{\text{ST}_\text{val}}$. 
Moreover, VOC now allows e-e interactions to have an effect on $E_{\text{ST}_\text{val}}$.
We therefore find that $E_{\text{ST}_\text{val}}<E_\text{val}$, for the same reasons that $E_{\text{ST}_\text{orb}}<E_\text{orb}$ in the presence of e-e interactions.
On the other hand, VOC has a more mixed effect on $E_{\text{ST}_\text{orb}}$, whose behavior is nonmonotonic near $x_s\approx 0$.
For the larger dot shown in Fig.~\ref{fig:fig4L}(d), $E_\text{orb}<E_\text{val}$ and $E_{\text{ST}_\text{orb}}<E_{\text{ST}_\text{val}}$ in the asymptotic regime, $|x_s|\gg 0$.
In this case, the $\mathrm{T_{orb}}$ and $\mathrm{T_{val}}$ levels anticross, and we find that their orbital and valley characters switch at the anticrossing.
Since e-e interactions are very strong for a large dot, $E_{\text{ST}_\text{val}}$ is suppressed very strongly below $E_\text{val}$. 
In all cases, we find that $E_\text{ST}$ is strongly suppressed near a step.

\subsubsection{Tilted interface}
Another common form of interface disorder occurs when the interface is tilted slightly away from the crystallographic axie; this can occur, for example, when the heterostructure is grown on a slightly miscut substrate. 
Here, we model such tilt by introducing uniformly distributed atomic steps at the interface. 
As for the case of a single step, the dot-step separation, $W$, plays an important role in determining the VOC. 

In the present case, the dot may overlap with several steps, so the dot diameter $D = 2\sqrt{\hbar/m_t \omega}$ [depicted in the inset of Fig.~\ref{fig:fig3L}(d)] also plays a role in determining the VOC. 
More specifically, we can expect the ratio $W/D$ to have a strong effect on both the single-electron valley splitting and the two-electron splitting $E_{\text{ST}_\text{val}}$.
We explore this effect in Fig.~\ref{fig:fig3L}(d) for the case where $d=W/2$; the other system parameters are the same as Fig.~\ref{fig:fig3L}(a) here, with $e^*=e$.
When $W\gg D$, we recover the excitation energies of Fig.~\ref{fig:fig3L}(a).
When $W\lesssim D$, the valley splitting is suppressed and would eventually go to zero in the limit $W\ll D$.
As in Fig.~\ref{fig:fig4L}, we find that $E_{\text{ST}_\text{val}}<E_\text{val}$, and that $E_{\text{ST}_\text{orb}}$ is only weakly affected by VOC.

\begin{figure}
\includegraphics[width=3.4 in]{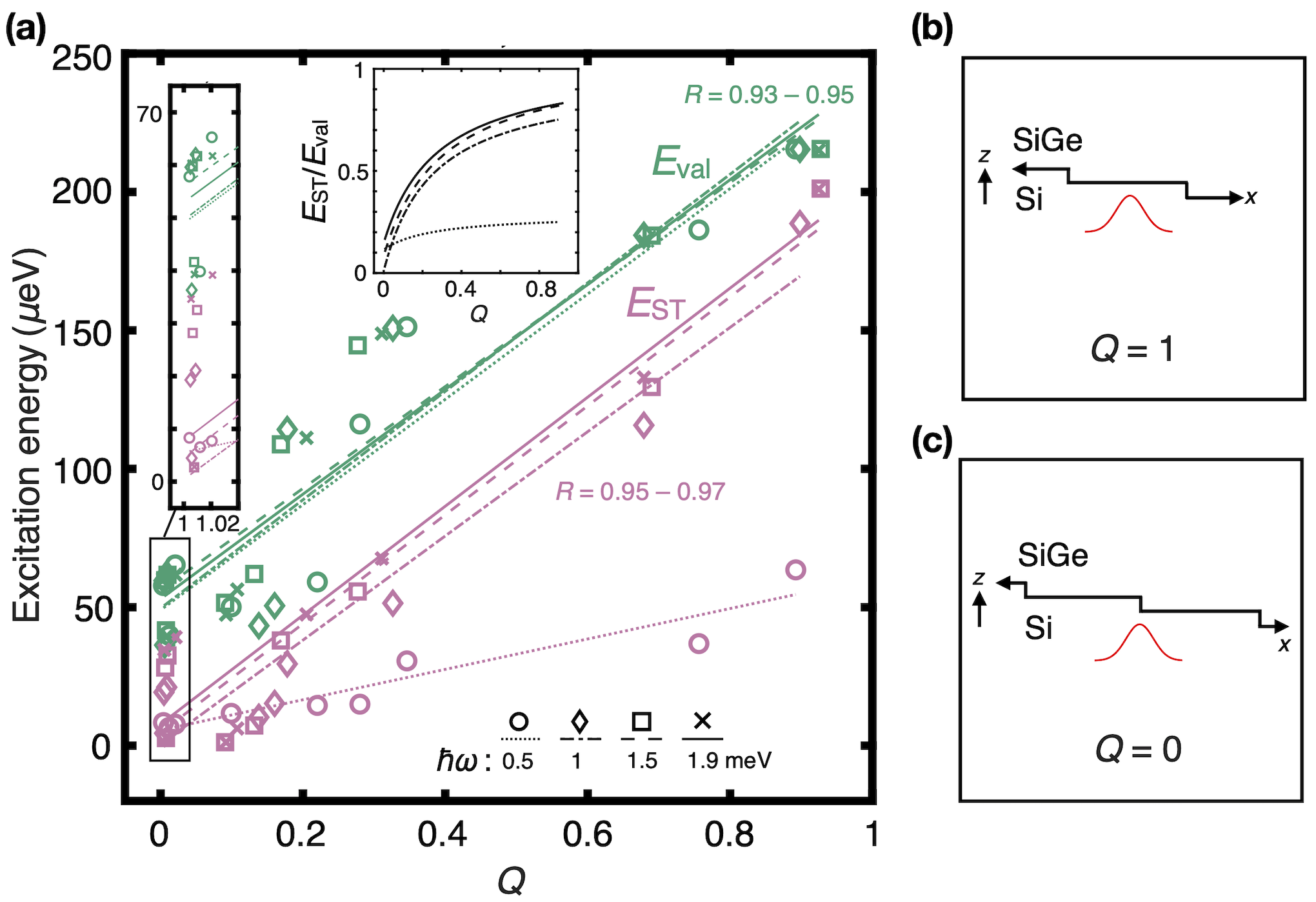}
\caption{One and two-electron excitation energies for dots at a tilted interface, for a range step widths, dot diameters, and dot-step separations.
[See inset of Fig.~\ref{fig:fig3L}(d) for illustration.]
(a) Results are combined using a single parameter $Q$, defined in Eq.~(\ref{eq:Qdef}), which characterizes the localization of an electron with respect to a step (or steps), as illustrated in (b) and (c).
$E_\text{ST}$ exhibits two distinct slopes as a function of $Q$, depending on its T$_\text{orb}$ or T$_\text{val}$ character, with $E_\text{ST}\rightarrow 0$ ($Q\rightarrow 0$), corresponding to strong VOC.
For all calculations, we assume a circular dot in a 10~nm wide quantum well, with an electric field of 1.5~MV/m and $\hbar\omega$ as indicated in the figure.
}
\label{fig:fig5L}
\end{figure}

In a companion paper \cite{Ercanprep}, we showed that the approximate valley symmetry leading to $E_{\text{ST}_\text{val}}\approx E_\text{val}$, also extends to cases with $W\gtrsim D$, for a tilted interface.
In the present work, we consider the full interplay between parameters $W$, $D$, and $d$.
To begin, we compute the two-electron energy excitations for all possible combinations of the following step parameters: $W/D \in \{3/5,1,2\}$, $d/W \in \{0,1/4,1/2\}$, and $\hbar \omega / \mathrm{meV} \in \{0.5,1,1.5,1.9\}$.
These results are reported in Appendix \ref{sec:app3}.
Here, we combine the results into a more cohesive form, based on the following observation:
although the valley phases differ by $\sim$$\pi$ for a single-atom step~\cite{Friesen:2007p115318} (i.e., nearly maximally), step heights of two atoms have a much weaker effect on the valley phase~\cite{Friesen:2010p115324}.
We can account for this even/odd effect by defining a parameter
\begin{equation}
Q=\left|1-2\sum_{i = \mathrm{odd}}\int_{s_i} |\Psi_\text{S}(\mathbf{r}_1,\mathbf{r}_2)|^2 d\mathbf{r}_1d\mathbf{r}_2\right| ,
\label{eq:Qdef}
\end{equation}
where the integral is performed over a given step $s_i$, and the sum is taken over every other step.
The extremal values of $Q$ are illustrated in Figs.~\ref{fig:fig5L}(b) and \ref{fig:fig5L}(c) for strongly localized dot wavefunctions.
The function $Q$ can be used to characterize a single-step interface or a many-step interface; it can also be adapted for one-electron or many-electron systems.
In all cases, $Q=0$ indicates strong VOC and $Q=1$ indicates weak VOC. 

In Fig.~\ref{fig:fig5L} we plot $E_\text{ST}$ and $E_\text{val}$ as a function of $Q$,
obtaining results that collapse fairly well onto straight lines (note the large $R$ values), with slopes that depend on the confinement $\hbar\omega$.
The data collapse suggests that Eq.~(\ref{eq:Qdef}) indeed describes the effect of VOC on both one- and two-electron excitation energies.
The collapse is poorer for $E_{\mathrm{val}}$ (cyan data), as expected, because $E_{\mathrm{val}}$ is a single-electron quantity, whereas $Q$ [as defined in Eq.~(\ref{eq:Qdef})] and $E_\text{ST}$ are two-electron quantities. 

The two $E_\text{ST}$ slopes in Fig.~\ref{fig:fig5L} correspond to the different excitation regimes in Fig.~\ref{fig:fig3L}(b): $E_{\text{ST}}=E_{\mathrm{ST_{orb}}}$ 
for weak confinement ($\hbar\omega<1.1$~meV) or $E_{\text{ST}}=E_{\mathrm{ST_{val}}}$  for strong confinement ($\hbar\omega>1.1$~meV).
Here, the dot with $\hbar\omega=1$~meV sits near the crossover, and behaves similarly to the strongly confined dots.
When T$_\text{val}$ is the dominant triplet, $E_\text{ST}\rightarrow 0$ when $Q\rightarrow 0$, because $E_\text{val}$ is strongly suppressed near a step, similar to Fig.~\ref{fig:fig4L}(c).
When T$_\text{orb}$ dominates, $E_\text{ST}\rightarrow 0$ for the same reason, because the low-energy triplet takes on a strong valley-like character, similar to Fig.~\ref{fig:fig4L}(d).
The inset shows that $E_{\mathrm{ST}}$ goes to zero much faster than $E_\text{val}$, as consistent with Fig.~\ref{fig:fig3L}(d), due to the fact that the broken valley symmetry allows e-e interactions to modify and suppress $E_\text{ST}$.

While the behavior embodied in Eq.~(\ref{eq:Qdef}) is useful as a guide, it does not fully describe the range of behaviors that are observed for strongly interacting electrons at disordered interfaces. 
For example, we observe some cases with low $Q$ values but relatively high $E_{\mathrm{ST}}$ values, up to $\sim 40$~$\mu$eV (see the left-hand inset). 
Such outliers are infrequent, however, and do not spoil the general correlations observed in Fig.~\ref{fig:fig5L}; they also do not correspond to ST splittings that are large enough to be practical for quantum computing.

The main take-home message from this VOC analysis is that interfacial steps suppress $E_{\mathrm{ST}}$ and are therefore detrimental for quantum computing.
To take a specific example, the qubit energy should be much larger than the thermal energy, for initialization or readout purposes.
For electrons in Si/SiGe devices, this typically corresponds to 100~mK, for which $k_BT\approx 8.6$~$\mu$eV~\cite{Takeda:2013p123113,Thorgrimsson:2017p32,Dodson:2020p505001}.
To be safe, qubits therefore require $E_\text{ST}>80$~$\mu$eV.
To achieve such values, Fig.~\ref{fig:fig5L} suggests that we need $Q>0.4$, indicating that at least 70\% of an electron's probability should be centered on a single terrace (or on terraces with step-height differences of two atomic layers).
Moreover, the single-electron valley splitting should be at least $\sim$150~$\mu$eV.

\subsection{\label{subsec:elliptical_dots}Elliptical dots}   
We now examine the effects of anisotropy in the 2D confinement potential on the singlet-triplet splittings. 
We find that changing the confinement potential from circular to elliptical affects our results quantitatively, but not qualitatively.

\begin{figure}
\includegraphics[width=3.4 in]{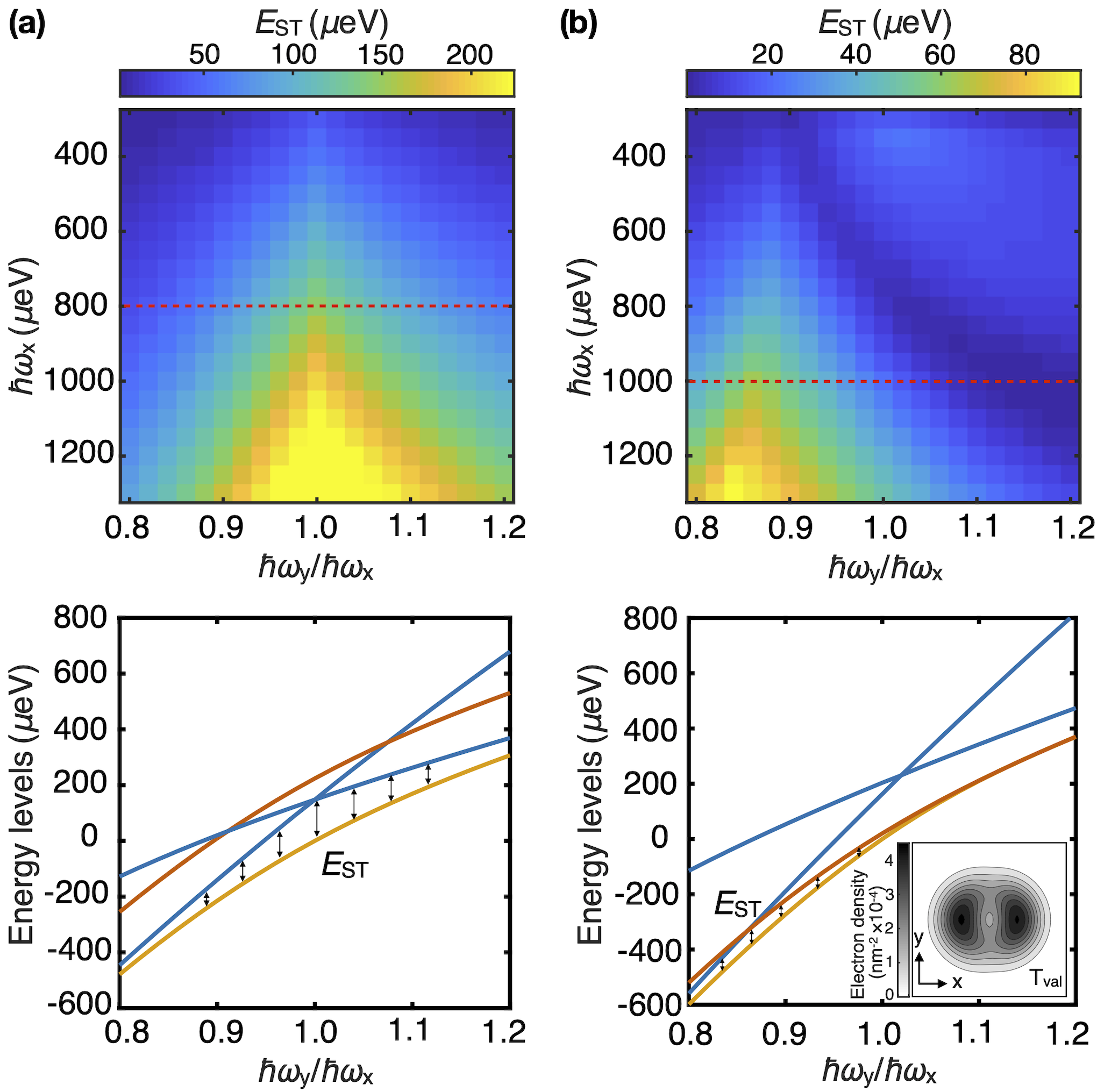}
\caption{Effects of anisotropic confinement, with and without interface steps.
All calculations assume quantum wells of width 9.1~nm, and a vertical electric field of 0.6~MV/m. 
Top panels show the dependence of $E_{\mathrm{ST}}$ on the 2D confinement parameters $\hbar \omega_x$ and $\hbar \omega_y/\hbar \omega_x$ for the case of (a) a flat interface, (b) a $0.23^{\circ}$ tilted interface (toward $\hat x$), with a step separation of $W\approx 34$~nm, and a dot-step separation of $d\approx 0.4W$. 
Bottom panels show the energies of the singlet (yellow), the valley triplet (red), and the two orbital triplets (blue), as a function of $\hbar \omega_y/\hbar \omega_x$, for the value of $\hbar \omega_x$ indicated by red-dashed lines in the top panels.
(a) $E_{\mathrm{ST}}$ peaks at the symmetry point $\hbar \omega_y=\hbar \omega_x$. 
Away from this point, the circular symmetry is lifted, and the T$_\text{orb}$ levels become non-degenerate.
(b) $E_{\mathrm{ST}}$ peaks at $\hbar \omega_y / \hbar \omega_x < 1$, due to the combination of confinement- and step-induced anisotropy. 
$E_\text{ST}$ is small compared to (a), due to the combination of VOC and e-e interactions. 
Inset: two-electron density of the low-energy triplet at point $\hbar\omega_x=\hbar\omega_y=800$~$\mu$eV when $d=0$, showing the effect of step-induced anisotropy.
}
\label{fig:fig6L}
\end{figure}

Once again, we begin with a flat interface, with no steps.
We assume a parabolic confinement potential, as before, but we allow for separate confinement strengths, $\hbar\omega_x$ and $\hbar\omega_y$, along $\hat x$ and $\hat y$.
In Fig.~\ref{fig:fig6L}(a), we plot $E_{\mathrm{ST}}$ as a function of both $\hbar \omega_x$ and the anisotropy ratio, $\hbar \omega_y / \hbar \omega_x$.
For any given $\hbar \omega_x$, $E_{\mathrm{ST}}$ is maximized when the confinement is isotropic, $\hbar \omega_y = \hbar \omega_x$.
This is easy to understand because breaking the circular symmetry causes the degeneracy of the two-electron orbital triplets to be lifted, as shown in the lower panel of Fig.~\ref{fig:fig6L}(a) (blue curves).
The singlet-triplet splitting, $E_\text{ST}$, is the energy difference between the lowest of these two curves and the singlet curve (yellow), and the peak in $E_{\mathrm{ST}}$ occurs where these triplet states cross in energy.
For the confinement energy $\hbar\omega_x=800$~$\mu$eV assumed in this plot, indicated by the dashed line in the top panel, T$_\text{orb}$ is always the dominant excitation.

Next, we add a uniform tilt to the interface to induce VOC, with results shown in Fig.~\ref{fig:fig6L}(b).
As discussed in previous sections, the combination of VOC and strong e-e interactions suppresses the overall magnitude of $E_{\mathrm{ST_{val}}}$. 
The steps also break circular symmetry, as seen in Fig.~\ref{fig:fig3L}(d), so there are two sources of anisotropy.
The bottom panel of Fig.~\ref{fig:fig6L}(b) shows the two-electron energy levels for $\hbar\omega_x=1000$~$\mu$eV.
Here, T$_\text{val}$ (red curve) is suppressed below $\mathrm{T_{orb}}$ over most of the anisotropy range of interest, making T$_\text{val}$ the dominant excitation.
$E_{\text{T}_\text{val}}$ crosses the energy level of one of the orbital triplets at around  $\hbar \omega_y / \hbar \omega_x = 0.85$, causing the peak in $E_{\mathrm{ST}}$. 

In Fig.~\ref{fig:fig6L}(b), we see that the peak in $E_{\mathrm{ST}}$ shifts to the left, as a consequence of having two different anisotropy sources.
To understand this, we first note that when $\hbar\omega_x<\hbar\omega_y$, and when there are no steps, the first orbitally excited single-electron state has a node in the $\hat x$ direction.
The structure along $\hat x$ for single-electron wavefunctions also extends to two-electron wavefunctions; however, selection rules for the Coulomb matrix elements prevent excitations from occurring in the $\hat y$ direction for the two-electron wavefunctions.
The key observation here is that elongation of the wavefunction along $\hat x$ occurs on the right-hand side of Fig.~\ref{fig:fig6L}(a).
Now, when we include a single step in the $x$-$z$ plane, but remove all other anisotropy, our FCI calculations show that the emerging Wigner-molecule structure also occurs in the $\hat x$ direction, as shown in the inset of Fig.~\ref{fig:fig6L}(b).
Here, the key observation is that this behavior is similar to what we already found on the right-hand side of Fig.~\ref{fig:fig6L}(a).
Hence, the step acts like an effective, built-in softening of the $x$ confinement, which results in the peak in Fig.~\ref{fig:fig6L}(b) shifting to the left.
In this case, we note again that selection rules prevent excitations from occurring along the $\hat y$ direction, for two-electron wavefunctions.

\section{\label{sec:conclusion}Summary and Conclusions}
In this paper we have studied the combined effects of strong electron-electron interactions and interface disorder in two-electron quantum dots in silicon heterostructures.
Our calculations combine
full configuration interaction and tight-binding methods to treat interaction effects and valley-orbit coupling nonperturbatively and consistently.

Electron-electron interactions become more important as the dot size increases.
Our calculations show that for dots in Si/SiGe quantum wells, in the experimentally relevant confinement regime (with typical dot sizes of several tens of nanometers), electron-electron interactions are strong. 
For a smooth quantum well interface, without disorder, the valley degree of freedom does not couple to the orbital degree of freedom.
Two-electron excitations therefore take the form of well-defined and qualitatively distinct valley triplets or orbital triplets.
Interactions have a strong effect on the energy of these two-electron states, but for typical dot sizes, the singlet-to-valley-triplet excitation energy is very close to the single-electron valley-splitting energy, even when interactions modify the shape of the valley triplets substantially.

Atomic steps at the quantum well interface are known to suppress the valley splitting. 
They are the main source of valley-orbit coupling, which hybridizes the single-electron valley and orbital eigenstates, and allows strong electron-electron interactions to suppress the valley-triplet excitation energy, relative to the valley splitting.
Steps also reduce the confinement symmetry in a dot, effectively reducing the confinement in the direction perpendicular to the steps.
The step density, and the position of step(s) relative to the dot, both have a strong effect on the singlet-triplet splitting. 
Importantly, the observation of a large singlet-triplet splitting is a good indictor of a large valley splitting, and suggests that atomic steps (if present) do not have a strong effect on the dot.

Finally, we note that the results described here also pertain to qubits formed in MOS systems, such as Si/SiO$_2$ interfaces or Si finFETs~\cite{Veldhorst:2014p981,Jock:2018p42000,Leon:2020p797,Yang:2020p350,Petit:2020p355,Bosco:2021p010348,Camenzid:preprint}. 
However, for those systems, the electric fields are typically higher than in SiGe/Si/SiGe quantum wells, which enhances the valley splitting as well as confinement energies.
As a result, electrons in MOS dots are less likely to be strongly interacting.

In conclusion, we have presented calculations showing that electron-electron interactions are significant in multi-electron Si/SiGe quantum dots. 
We have also shown that valley physics is nontrivial in these devices, particularly for strongly interacting electrons, with implications for the energy spectrum.
Interfacial disorder adds an extra layer of complexity, as it couples the valley and orbital degrees of freedom. 
The current study provides key insights into the physics of multi-electron quantum dots in silicon, which is essential for the rational design and operation of qubits.

\begin{acknowledgments}
The authors thank Mark Eriksson, J.\ P.\ Dodson, Joelle Corrigan, Tom McJunkin, Leah Tom, Merritt Losert, Jos\'{e} Carlos Abadillo-Uriel, Mark Gyure, Samuel Quinn, Andrew Pan and Joseph Kerckhoff for helpful discussions. This work was supported in part by ARO through Award No.\ W911NF-17-1-0274 and the Vannevar Bush Faculty Fellowship program sponsored by the Basic Research Office of the Assistant Secretary of Defense for Research and Engineering and funded by the Office of Naval Research through Grant No.\ N00014-15-1-0029. The views and conclusions contained in this document are those of the authors and should not be interpreted as representing the official policies, either expressed or implied, of the U.S. Government. The U.S. Government is authorized to reproduce and distribute reprints for Government purposes notwithstanding any copyright notation herein. This research was performed using the compute resources and assistance of the UW-Madison Center For High Throughput Computing (CHTC) in the Department of Computer Sciences. The CHTC is supported by UW-Madison, the Advanced Computing Initiative, the Wisconsin Alumni Research Foundation, the Wisconsin Institutes for Discovery, and the National Science Foundation, and is an active member of the Open Science Grid, which is supported by the National Science Foundation and the U.S. Department of Energy's Office of Science.
\end{acknowledgments}
\appendix

\section{\label{sec:app1}Overview of the calculation strategy}
\begin{figure*}
\includegraphics[width=1 \textwidth]{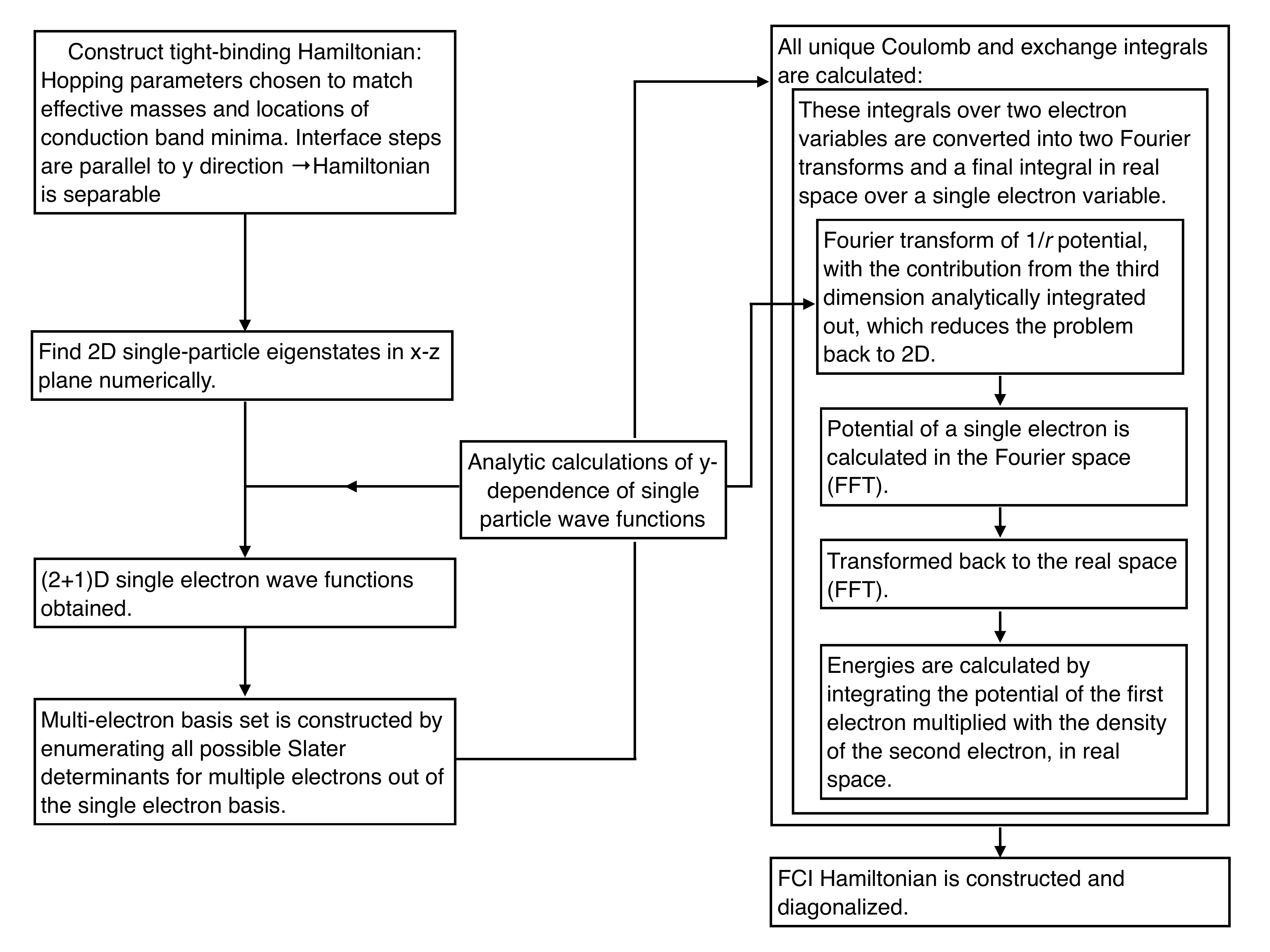}
\caption{A flowchart summary of the methods used in our calculations, combining tight-binding calculations that incorporate atomic-scale disorder with full configuration interaction calculations to account for strong electron-electron interactions, nonperturbatively.
}
\label{fig:A1L}
\end{figure*}

This Appendix provides an overview of the calculational methods.
The information is summarized in
Fig.\ \ref{fig:A1L}, which illustrates how the
tight-binding and the configuration-interaction methods are integrated.
\\

\section{\label{sec:app1b}Details related to computing the Coulomb matrix elements}
This Appendix presents details related to the evaluation of
Coulomb matrix elements of the form
$\bra{\psi_{\alpha}} H_{\mathrm{int}} \ket{\psi_{\beta}}$. The overall approach of using Fourier transforms to accelerate the calculations is adapted from Ref.~\cite{Molnar:2002p7795}.

After the spin degrees are summed over, the integrals to be performed are of the form
\begin{equation}
 (i \ j | k \ l)\equiv\int d\bm{r}_1d\bm{r}_2\phi_i^*(\bm{r}_1)\phi_j(\bm{r}_1)H_{\mathrm{int}} \phi_k^*(\bm{r}_2)\phi_l(\bm{r}_2).
 \label{mainIntegral}
\end{equation} 
Since the energy eigenstates are separable, $\phi(x,y,z)=\zeta(x,z)\eta(y)$, the integral $(i \ j | k \ l)$ becomes
\begin{widetext}
\begin{equation} 
 \frac{e^2}{4 \pi \epsilon_0 \epsilon_r } \int dx_1dz_1dx_2dz_2\zeta_{i_{xz}}^*(x_1,z_1)\zeta_{j_{xz}}(x_1,z_1)\zeta_{k_{xz}}^*(x_2,z_2)\zeta_{l_{xz}}(x_2,z_2)
  f_{ijkl}(\Delta r_\perp), \hfil
\end{equation}
  where $f_{ijkl}(\Delta r_\perp) \equiv \int dy_1dy_2\frac{\eta_{i_y}^*(y_1)\eta_{j_y}(y_1)\eta_{k_y}^*(y_2)\eta_{l_y}(y_2)}{\sqrt{(y_1-y_2)^2+\Delta r_\perp^2}}$ and $\Delta r_\perp^2\equiv(x_1-x_2)^2+(z_1-z_2)^2$.
  
For the harmonic confinement potentials assumed here,
the wavefunctions $\eta(y)$ are harmonic oscillator eigenstates, given by
 \begin{equation}
 \eta_n(y)=\left(\frac{1}{\pi L^2}\right)^{1/4}\frac{1}{\sqrt{2^n n!}}H_n (y/L)e^{-y^2/2L^2},
\hspace{1in}
L\equiv\sqrt{\hbar/m^*\omega}, 
 \end{equation}
 where $H_n(y/L)$ are the Hermite polynomials.
 Hence, the $y$-integral takes the form
 \begin{equation}
 f_{ijkl}(\Delta r_\perp)=\sum\limits_{n,m} c_{nm}\int dy_1dy_2\frac{y_1^n y_2^m e^{-(y_1^2 + y_2^2)/L^2}}{\sqrt{(y_1-y_2)^2 + \Delta r_\perp^2}}.
 \end{equation}
 Now, we make the change of variables $s\equiv y_1-y_2$ and $t\equiv y_1+y_2 $, so that
 \begin{equation}
 f_{ijkl}(\Delta r_\perp)=\sum\limits_{n,m} \frac{c_{nm}}{2^{m+n+1}} \sum\limits_{k,p=0}^{n,m} \binom{n}{k} \binom{m}{p} (-1)^p \int ds \frac{s^{n+p-k} e^{-s^2/2L^2}}{\sqrt{s^2 + \Delta r_\perp^2}} \int dt \ t^{k+m-p} e^{-t^2/2L^2}.
\end{equation} 
The integrals inside the summation have the following analytical forms: \begin{subequations}
\begin{equation}
\int ds \frac{s^q e^{-s^2/2L^2}}{\sqrt{s^2 + \Delta r_\perp^2}}= 
 \begin{cases}
\sqrt{\pi} L^q (q-1)!! \ U(1/2,\frac{2-q}{2},\Delta r_\perp^2/2L^2), & \text{if } q=\text{even} \\
0, & \text{otherwise}
\end{cases} 
 \end{equation}

\begin{equation}
 \int dt \ t^q e^{-t^2/2L^2}= 
  \begin{cases}
L^{q+1} \Gamma(\frac{q+1}{2}) 2^{(q+1)/2}, & \text{if } q=\text{even} \\
0, & \text{otherwise,}
\end{cases} 
\end{equation}
\label{Tricomi}
 \end{subequations}
where $U$ is the Kummer function~\cite[Eq.~13.4.4,13.2.40]{nist} and $\Gamma$ is the factorial function~\cite[Eq.~5.2.1]{nist}.

We define
\begin{equation}
\begin{aligned}
&\rho^{(1)}(\bm{r}_1)\equiv\phi_i^*(\bm{r}_1)\phi_j(\bm{r}_1), \\ &\rho^{(2)}(\bm{r}_2)\equiv\phi_k^*(\bm{r}_2)\phi_l(\bm{r}_2),\\ &\text{and} \ V(\bm{r}_2)=\frac{e}{4 \pi \epsilon_0 \epsilon_r}\int d\bm{r}_1 \frac{\rho^{(1)}(\bm{r}_1)}{\abs{\bm{r}_1-\bm{r}_2}}.
\end{aligned}
\end{equation}
(We will also use $\rho^{(1,2)}(x_{1,2},z_{1,2})=\zeta^*_{i_{xz}}(x_{1,2},z_{1,2}) \zeta_{j_{xz}}(x_{1,2},z_{1,2})$, and $\rho^{(1,2)}(y_{1,2})=\eta^*_{i_{y}}(y_{1,2}) \eta_{j_{y}}(y_{1,2})$, for simplicity of notation.) A general matrix element then has the form $I=\int dr_2 \rho^{(2)}(r_2) V(r_2)$, where the subscripts $ijkl$ have been dropped for convenience. 
We integrate out the $y$ terms and apply the forms defined in~Eq.~(\ref{Tricomi}): 
\begin{equation}
\begin{aligned}
V(x_2,z_2) &= \int dy_2 V(\bm{r}_2) \\ 
&= \frac{e}{4 \pi \epsilon_0 \epsilon_r} \int d\bm{r}_1 dy_2 \frac{\rho^{(1)}(\bm{r}_1)}{\abs{\bm{r}_1-\bm{r}_2}} \\
&=\frac{e}{4 \pi \epsilon_0 \epsilon_r} \int dx_1 dz_1 \rho^{(1)}(x_1,z_1) f(\Delta r_\perp).
\end{aligned}
\end{equation}

The problem is reduced to 2D by replacing the $1/r$ potential with the effective potential, $f$. Now, the Fourier integral representations of $f$ and $\rho$ are
\begin{subequations}
\begin{equation} 
\rho^{(1)}(x_1,z_1)=\int d\bm{q} e^{2 \pi i\bm{q} \cdot \bm{r}_{1\perp}} \tilde{\rho}^{(1)}(q_x,q_z)
\end{equation}
\begin{equation}
f(\Delta r_\perp)=\int d\bm{k} e^{2 \pi i\bm{k} \cdot (\bm{r}_{1\perp}-\bm{r}_{2\perp})} \tilde{f}(k_{x},k_{z})
\end{equation}
\end{subequations}
Using these forms and completing the spatial integrals, $V$ can also be expressed as a Fourier transform:
\begin{equation}
\begin{aligned}
\frac{4 \pi \epsilon_0 \epsilon_r}{e} V(x_2,z_2)&= \int dx_1 dz_1 \int d\bm{k} e^{2 \pi i\bm{k} \cdot (\bm{r}_{1\perp}-\bm{r}_{2\perp})} \tilde{f}(k_{x},k_{z})
\int d\bm{q} e^{2 \pi i\bm{q} \cdot \bm{r}_{1\perp}} \tilde{\rho}^{(1)}(q_x,q_z) \hfill \\
&= \int dx_1 dz_1 e^{2 \pi i(\bm{k}+\bm{q}) \cdot \bm{r}_{1\perp}}
\int d\bm{q} d\bm{k} \tilde{f}(k_{x},k_{z}) \tilde{\rho}^{(1)}(q_x,q_z) e^{-2 \pi i\bm{k} \cdot \bm{r}_{2\perp}} \hfill \\
&= \int d\bm{q} d\bm{k} \delta(q_{x}+k_{x}) \delta(q_{z}+k_{z}) \tilde{f}(k_{x},k_{z})
\tilde{\rho}^{(1)}(q_x,q_z) e^{-2 \pi i\bm{k} \cdot \bm{r}_{2\perp}} \hfill \\
&= \int d\bm{k} e^{2 \pi i\bm{k} \cdot \bm{r}_{2\perp}} \tilde{f}(k_{x},k_{z}) \tilde{\rho}^{(1)}(k_x,k_z) \hfill
\end{aligned}
\label{Eq:potential}
\end{equation}
The final integration in Eq.~(\ref{Eq:potential}) is performed using the \textsc{matlab} FFT routine. $\tilde{\rho}^{(1)}$ is also obtained by the FFT routine on $\rho^{(1)}$, which is obtained using the TB method described in the main text. The Fourier transform of $f$ is performed analytically:
\begin{equation}
\tilde{f}(k_x,k_z)= \int_{0}^{D_c} dr~r \int_{0}^{2 \pi} d\theta e^{2 \pi ikr \cos\theta}f(r)= 2 \pi \int_{0}^{D_c} J_0(\abs{2 \pi kr})f(r) r dr \hfill .
\end{equation}
Here, $J_0$ is the 0th order Bessel function of the first kind~\cite[Eq.~10.9.1]{nist}, and $D_c$ is a cutoff distance that prevents the charges from interacting with their periodic images~\cite{Molnar:2002p7795}. Since the resulting expression for $f$ is a sum of confluent hypergeometric functions $U(a,b,z)$, one must perform integrals
 of the form 
\begin{equation}
\int_{0}^{D_c} dr \ r \ J_0(\abs{2 \pi kr}) \ U(1/2,b,r^2/2L^2) . 
\label{int_D}
\end{equation}
These integrals need to be calculated numerically for each value of $k$, which turns out to be computationally expensive. However, when $D_c$ goes to infinity, there is an analytical solution, given by~\cite[Eq.~13.10.15]{nist}
\begin{equation}
 \int_{0}^{\infty} dr \ r \ J_0(\abs{2 \pi kr}) \ U(1/2,b,r^2/2L^2)= \frac{L^2}{\sqrt{\pi}} \Gamma(2-b) \ U(2-b,3/2,2 \pi^2 k^2 L^2). \hfill
  \label{int_inf}
 \end{equation}
If the integral from $D_c$ to infinity can be calculated efficiently, it can be subtracted from \eqref{int_inf} to obtain the desired integral \eqref{int_D}. For large $z$, the asymptotic form of $U$ is given by
\begin{equation}
U(a,b,z) \sim z^{-a} \sum\limits_{n=0}^{\infty}\frac{(a)_n (a-b+1)_n}{n!}(-z)^{-n},
\label{Eq:kummer_asymp}
\end{equation}
where $(a)_n$ is the Pochhammer symbol, defined as $\frac{\Gamma(a+n)}{\Gamma(a)}$~\cite[Eq.~13.7.3]{nist}. For each term in the sum, the integral has the following analytical form:
\begin{multline}
\int_{D_c}^{\infty} dr \ r \ J_0(\abs{2 \pi kr}) \left( \frac{r^2}{2 L^2} \right)^{-\frac{1}{2}-n} \\ = 2^{n+1/2} L^{2n+1} \left( \frac{4^{-n} (2 \pi k)^{2n-1} \Gamma(1/2-n)}{\Gamma(1/2+n)}  +\frac{D_c^{1-2n}}{2n-1} {}_1F_2(1/2-n;1,3/2-n;-\pi^2 D_c^2k^2 \right), 
\end{multline}
where ${}_pF_q$ is the generalized hypergeometric function~\cite[Eq.~16.2.1]{nist}.  We truncate the sum in Eq.~(\ref{Eq:kummer_asymp}) at the $n=5$ term; for the smallest value of the argument used in this work ($D_c^2/2L^2=38.87$) the resulting error is less than $0.2 \%$.

The discrete lattice in the $x$-$z$ plane has a maximum wavevector. Therefore, it is important to show that the $k$-space representations of the integrands fall off quickly enough to converge properly. In Fig.~\ref{fig:A2L}, we show that the expression $\frac{e}{4 \pi \epsilon_0 \epsilon_r}\tilde{f}(k_{x},k_{z}) \tilde{\rho}^{(1)}(k_x,k_z)$, in the integrand of Eq.~(\ref{Eq:potential}), indeed falls off quickly, for the case of  $\hbar \omega=1.9$~meV in Fig.\ \ref{fig:fig3L}(b). We choose to demonstrate a strongly-confined case here, since functions that are more localized in the real space are less localized in $k$-space. Note that the actual $k$-space domain is significantly larger than shown in these plots, ensuring the maximum wavevectors set by the discrete lattice are large enough.

\begin{figure}
\includegraphics[width=3.5 in]{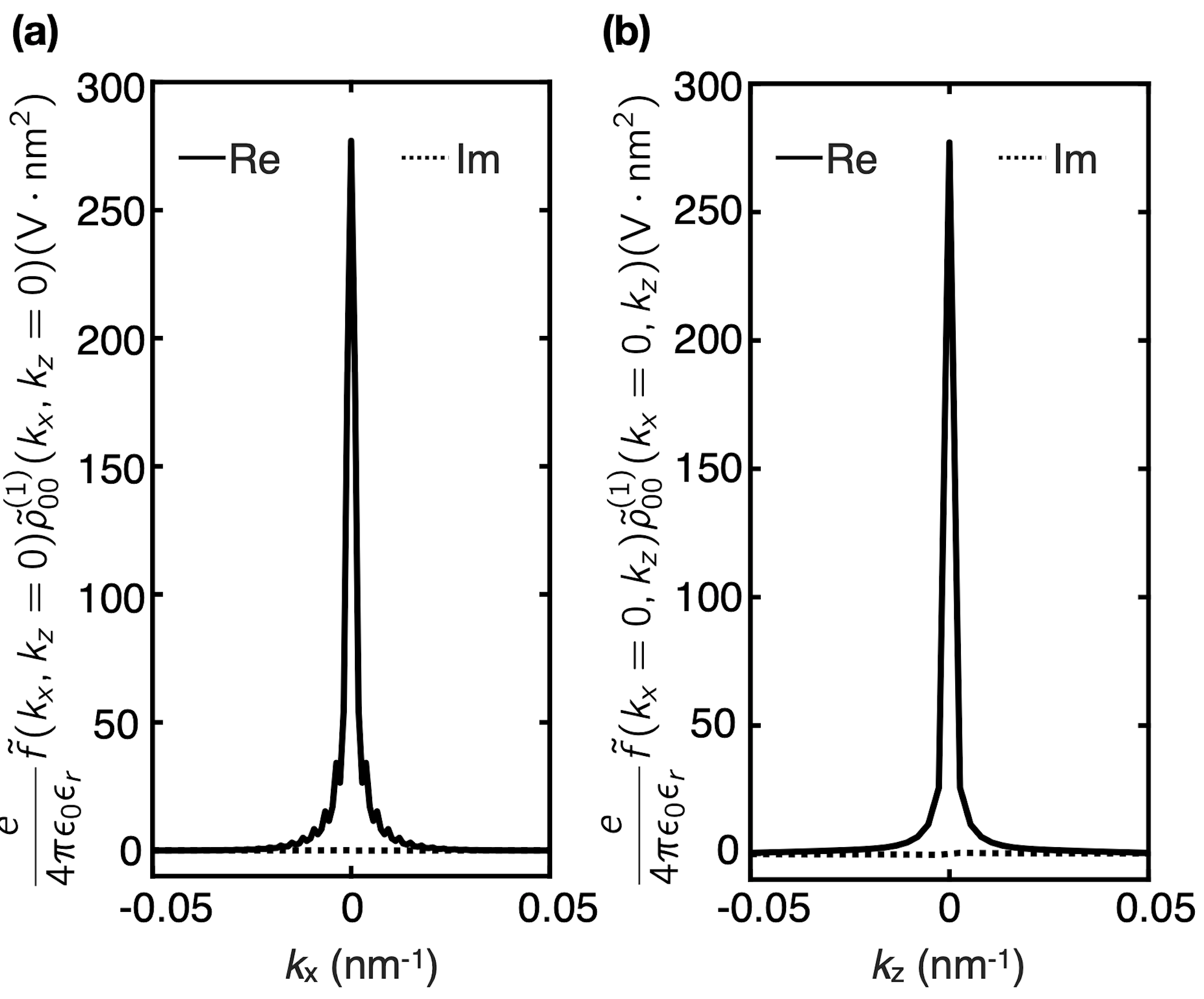}
\caption{$k$-space representation of the integrand in Eq.~(\ref{Eq:potential}). Here, $\rho_{00}$ is the single-electron ground state density for the $\hbar\omega=1.9$~meV data point in Fig.~\ref{fig:fig3L}(b). Only the central portions of the full domains, with significant amplitudes, are shown. (a) max $|k_x|=0.18$ nm$^{-1}$, $|k_z|=0$, (b) max $|k_z|=3.68$ nm$^{-1}$, $|k_x|=0$.
}
\label{fig:A2L}
\end{figure}

\section{Detailed discussion of why the valley singlet-triplet splitting $E_{\mathrm{ST_{val}}}$ is close to the single-particle valley spitting $E_{\mathrm{val}}$ even when interactions are strong}\label{sec:app2}
Here, we more explicitly discuss why the one- and two-electron excitation energies $E_{\mathrm{val}}$ and  $E_{\mathrm{ST_{val}}}$ are nearly equal, regardless of the confinement strength, in the absence of VOC, as discussed in Sec.~\ref{sec:ST}. As mentioned in the main text, this is due to two important observations: (i) the interaction integrals involving electrons with two different valleys are small due to the fast oscillating factors in the wavefunctions, (ii) valley states with the same orbital number have approximately the same envelopes. Therefore, integrals like $(a \ b | a^\prime \ b )$, $(a^\prime \ b | a^\prime \ b )$, $(a \ b^\prime | a^\prime \ b )$ are small, and $(a \ b | c \ d) \approx (a \ b | c^\prime \ d^\prime) \approx (a^\prime \ b^\prime | c \ d) \approx (a^\prime \ b^\prime | c^\prime \ d^\prime)$. [See Eq.~(\ref{mainIntegral}) for notation definitions.] We demonstrate these two effects numerically in Fig.~\ref{fig:A3L}, where we show results of calculations for a large number of representative matrix elements.

\begin{figure*}
\includegraphics[width=7 in]{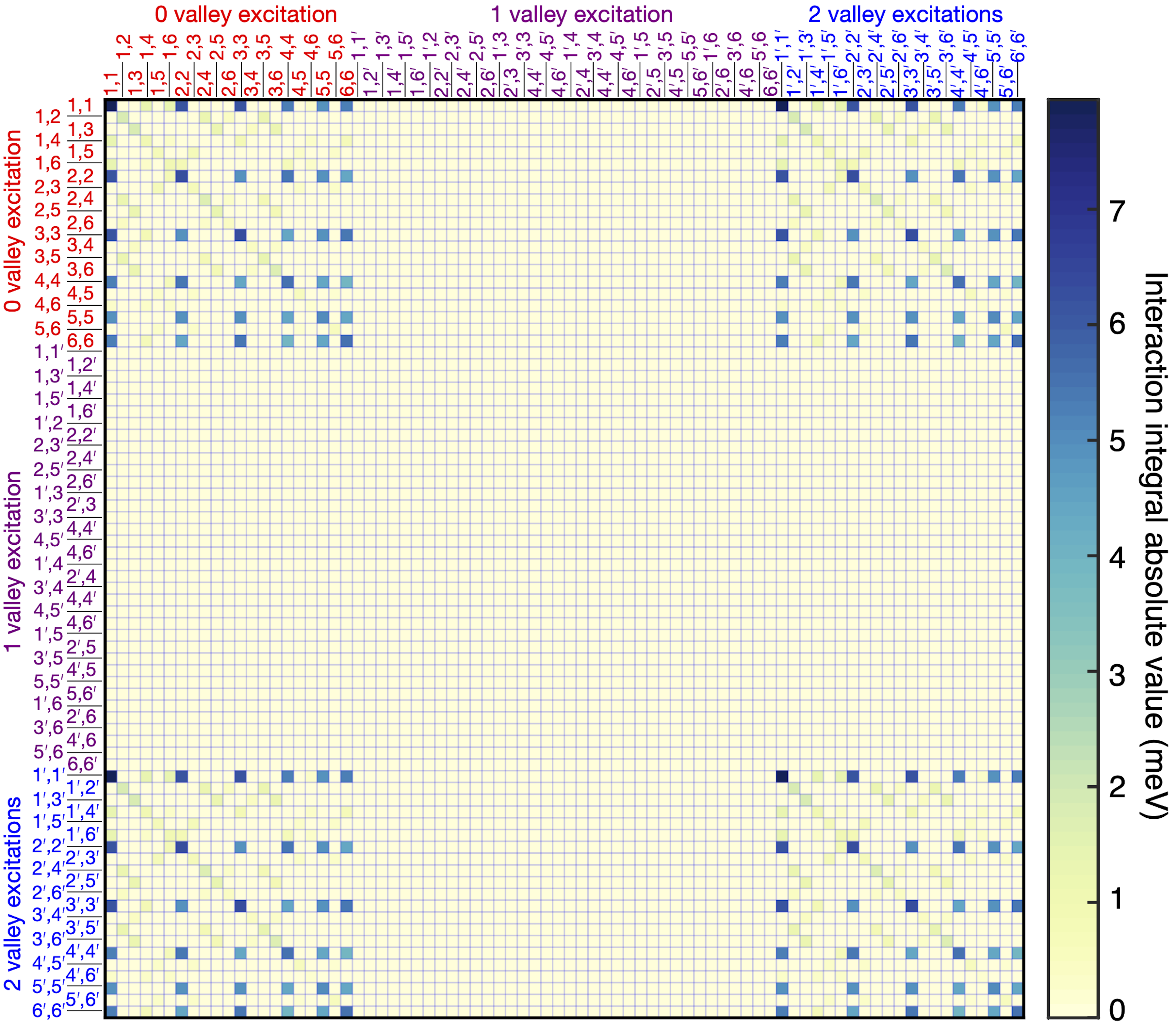}
\caption{Spatial interaction integrals involving states in the first three single-electron shells, corresponding to the case $\hbar \omega =1.2$~meV in Figs.~\ref{fig:fig3L}(b) and \ref{fig:fig3L}(c).}
\label{fig:A3L}
\end{figure*}

To illustrate further the symmetries and cancellations of the interaction terms in determining the energy separation of S and $\mathrm{T_{val}}$, we consider an example case motivated by the solutions shown in Figs.~\ref{fig:fig3L}(b)-\ref{fig:fig3L}(c) and Table \ref{tab:table1}. We write
\begin{subequations}
\begin{equation}
\braket{\mathrm{S}}{\bm{r}}=\alpha \phi_1 \phi_1 + \frac{\beta}{2} \left[ \phi_1 \phi_4 + \phi_4 \phi_1 + \phi_1 \phi_6 + \phi_6 \phi_1  \right]
\end{equation}
\begin{multline}  
\braket{\mathrm{T_{val}}}{\bm{r}}=\frac{\alpha}{\sqrt{2}} \left[\phi_1 \phi_{1^\prime} - \phi_{1^\prime} \phi_1 \right]  + \frac{\beta}{2\sqrt{2}} [ (\phi_1 \phi_{4^\prime} - \phi_{4^\prime} \phi_1) - (\phi_{1^\prime} \phi_4 - \phi_4 \phi_{1^\prime} ) + (\phi_1 \phi_{6^\prime} - \phi_{6^\prime} \phi_1) - (\phi_{1^\prime} \phi_6 - \phi_6 \phi_{1^\prime})],
\end{multline}
\end{subequations}
where $\alpha, \beta > 0$ and $\phi_i$ are real. Note that we have only included components I and III from Table \ref{tab:table1}, for simplicity. The energy difference due to the single-electron part of the Hamiltonian is equal to $E_{\mathrm{val}}$. We now show that the interaction contributions (approximately) cancel. Using the symmetries $(i j|k l)=(k l|i j)=(j i|k l)$ for real wavefunctions, we have:
\begin{multline}
\bra{\mathrm{S}}H_{int}\ket{\mathrm{S}}=\alpha^2 \iinteg{1}{1}{1}{1} + 2 \alpha \beta [\iinteg{1}{1}{1}{4}+\iinteg{1}{1}{1}{6}] + \frac{\beta^2}{2} [\iinteg{1}{1}{4}{4} + \iinteg{1}{4}{4}{1} + 2\iinteg{1}{1}{4}{6} + 2\iinteg{1}{6}{4}{1} \\ + \iinteg{1}{1}{6}{6} + \iinteg{1}{6}{6}{1} ]
\label{eq:es}
\end{multline}

\begin{figure*}[t]
\includegraphics[width=5.5 in]{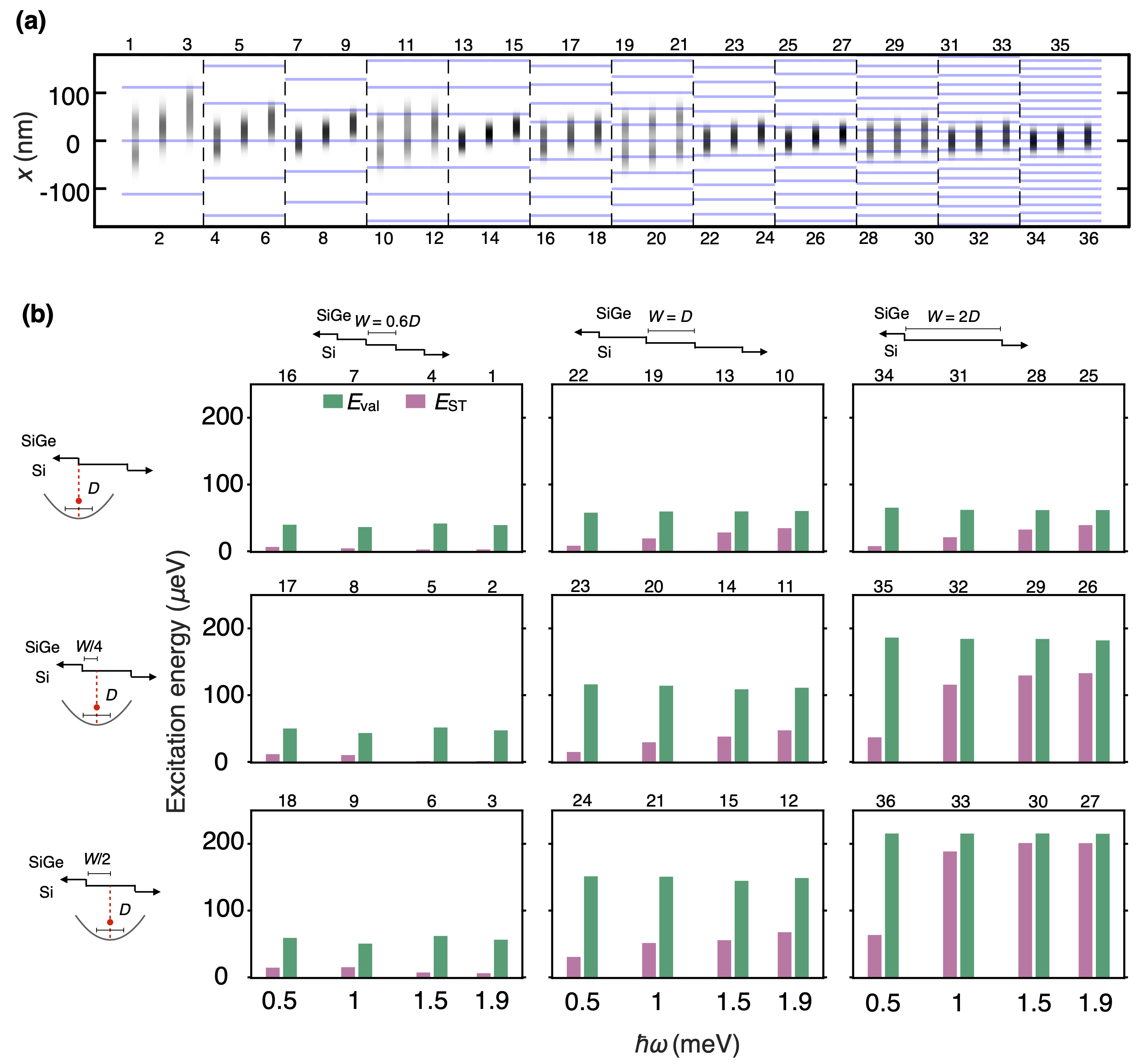}
\caption{Full set of solutions reported in Fig.~\ref{fig:fig5L}. (a) One-dimensional electron densities of the ground states, corresponding to 36 cases in Fig.~\ref{fig:fig5L}. Blue lines show the steps at the interface. (b) $E_{\mathrm{ST}}$ (pink) in comparison to $E_{\mathrm{val}}$ (cyan) in circular dots defined in 10~nm wide quantum wells, with an electric field of 1.5~MV/m and tilted interfaces. The columns show the horizontal distance between steps (the step widths, $W$), corresponding to 0.6$D$, $D$ and 2$D$, where $D=2\sqrt{\hbar/m_t \omega}$ is the dot diameter. The rows show the distance to the nearest step, $d$, corresponding to 0, $0.25W$, or $0.5W$. 
}
\label{fig:A4L}
\end{figure*}

\begin{table*}[t]
\caption{Values of some interaction integrals.} 
\begin{tabular}{c c c c c  c c c  c c}
\hline\hline
 \multicolumn{4}{c}{} &\hspace{.05in}& [Analytical: & Eqs.\ (\ref{eq:m1111})-(\ref{eq:m1212})] \hspace{.1in}&\hspace{.0in}& [Numerical: & Appendix \ref{sec:app1}] \\
 $l$  & $L$  & $\hbar \omega_z$ & $\hbar \omega_{xy}$ && $\iinteg{1}{1}{1}{1}$ & $\iinteg{1}{1^\prime}{1}{1^\prime}$ && $\iinteg{1}{1}{1}{1}$ & $\iinteg{1}{1^\prime}{1}{1^\prime}$  \\
(nm) & (nm) & (meV) & (meV) && (meV) & ($\mu$eV) && (meV) & ($\mu$eV) \\
 \colrule
 2.69 & 30.00 & 11.50 & 0.45 && 5.0 & 0.1 && 5.0 & 0.1 \\
 2.69 & 25.00 & 11.50 & 0.64 && 5.9 & 0.1 && 5.9 & 0.1 \\
 2.69 & 18.28 & 11.50 & 1.2 && 7.9 & 0.2 && 7.9 & 0.2 \\
 2.00 & 10.00 & 20.80 & 4.0 && 14.1 & 0.7 && 14.1 & 0.7 \\
 1.00 & 3.00 & 83.19 & 44.56 && 43.9 & 15.6 && 43.9 & 15.9 \\
 0.60 & 1.60 & 231.08 & 156.66 && 80.7 & 91.8 && 80.9 & 97.3 \\
\hline\hline
\end{tabular}
\label{tab:integrals}
\end{table*}

\begin{multline}
\bra{\mathrm{T_{val}}}H_{int}\ket{\mathrm{T_{val}}}=\alpha^2 [\iinteg{1}{1}{1^\prime}{1^\prime}-\iinteg{1}{1^\prime}{1^\prime}{1}] +\alpha\beta[\iinteg{1}{1}{1^\prime}{4^\prime}-\iinteg{1}{4^\prime}{1^\prime}{1}+\iinteg{1}{4}{1^\prime}{1^\prime}-\iinteg{1}{1^\prime}{1^\prime}{4} \\ + \iinteg{1}{1}{1^\prime}{6^\prime} - \iinteg{1}{6^\prime}{1^\prime}{1} + \iinteg{1}{6}{1^\prime}{1^\prime} - \iinteg{1}{1^\prime}{1^\prime}{6}] + \frac{\beta^2}{4}[\iinteg{1}{1}{4^\prime}{4^\prime}-\iinteg{1}{4^\prime}{4^\prime}{1}+\iinteg{1^\prime}{1^\prime}{4}{4}-\iinteg{1^\prime}{4}{4}{1^\prime} \\ +\iinteg{1}{1}{6^\prime}{6^\prime}-\iinteg{1}{6^\prime}{6^\prime}{1}+\iinteg{1^\prime}{1^\prime}{6}{6}-\iinteg{1^\prime}{6}{6}{1^\prime} + 2\iinteg{1}{1}{4^\prime}{6^\prime}-2\iinteg{1}{6^\prime}{4^\prime}{1}+2\iinteg{1^\prime}{1^\prime}{4}{6}-2\iinteg{1^\prime}{6}{4}{1^\prime} \\ +2\iinteg{1}{4}{4^\prime}{1^\prime}-2\iinteg{1}{1^\prime}{4^\prime}{4}+2\iinteg{1}{6}{4^\prime}{1^\prime}-2\iinteg{1}{1^\prime}{4^\prime}{6} +2\iinteg{1^\prime}{6^\prime}{4}{1}-2\iinteg{1^\prime}{1}{4}{6^\prime}+2\iinteg{1}{6}{6^\prime}{1^\prime}-2\iinteg{1}{1^\prime}{6^\prime}{6}] .
\end{multline}

Further simplifying, based on Fig.~\ref{fig:A3L} we obtain
\begin{multline}
\bra{\mathrm{T_{val}}}H_{int}\ket{\mathrm{T_{val}}} \approx \alpha^2 \iinteg{1}{1}{1^\prime}{1^\prime} +\alpha\beta[\iinteg{1}{1}{1^\prime}{4^\prime}+\iinteg{1}{4}{1^\prime}{1^\prime}+\iinteg{1}{1}{1^\prime}{6^\prime}+\iinteg{1}{6}{1^\prime}{1^\prime}] \\ + \frac{\beta^2}{4}[\iinteg{1}{1}{4^\prime}{4^\prime}+\iinteg{1^\prime}{1^\prime}{4}{4}+\iinteg{1}{1}{6^\prime}{6^\prime}+\iinteg{1^\prime}{1^\prime}{6}{6} + 2\iinteg{1}{1}{4^\prime}{6^\prime}+2\iinteg{1^\prime}{1^\prime}{4}{6}+2\iinteg{1}{4}{4^\prime}{1^\prime} \\ +2\iinteg{1}{6}{4^\prime}{1^\prime} +2\iinteg{1^\prime}{6^\prime}{4}{1}+2\iinteg{1}{6}{6^\prime}{1^\prime}] \\
= \alpha^2 \iinteg{1}{1}{1^\prime}{1^\prime} +2\alpha\beta[\iinteg{1}{1}{1^\prime}{4^\prime}+\iinteg{1}{1}{1^\prime}{6^\prime}] + \frac{\beta^2}{2}[\iinteg{1}{1}{4^\prime}{4^\prime}+\iinteg{1}{4}{4^\prime}{1^\prime}+2\iinteg{1}{1}{4^\prime}{6^\prime}+2\iinteg{1}{6}{4^\prime}{1^\prime} \\ +\iinteg{1}{1}{6^\prime}{6^\prime} +\iinteg{1}{6}{6^\prime}{1^\prime}]
\label{eq:etval}
\end{multline}
Comparing Eqs.~(\ref{eq:es})and(\ref{eq:etval}), we conclude that the interaction energies indeed cancel and, therefore, $E_{\mathrm{ST_{val}}} \approx E_{\mathrm{val}}$.

We now provide analytical estimates to back up these numerical observations. In particular, we compare $\iinteg{1}{1}{1}{1}$ and $\iinteg{1}{1^\prime}{1}{1^\prime}$  to show that the latter is much smaller than the former. Using the envelope function formalism \cite{Friesen:2007p115318}, with Gaussian envelopes for simplicity, we have
\begin{equation}
\phi_{1}=\frac{e^{-(x^2+y^2)/2L^2}}{(\pi L^2)^{1/2}} \frac{e^{-z^2/2l^2}}{(\pi l^2)^{1/4}} \sqrt{2}\sin{k_0 z}, \ \phi_{1^\prime}=\frac{e^{-(x^2+y^2)/2L^2}}{(\pi L^2)^{1/2}} \frac{e^{-z^2/2l^2}}{(\pi l^2)^{1/4}} \sqrt{2}\cos{k_0 z}.
\label{eq:env_wfs} 
\end{equation}

Therefore,
\begin{equation}
\iinteg{1}{1}{1}{1}=\frac{e^2}{4 \pi \epsilon_0 \epsilon_r } \int \frac{dz_1 dz_2}{\pi l^2}e^{-(z_1^2+z_2^2)/l^2} \sin^2{k_0z_1} \sin^2{k_0z_2} \int \frac{dx_1dy_1dx_2dy_2}{\pi^2 L^4} \frac{e^{-(x_1^2+y_1^2+x_2^2+y_2^2)/L^2}}{((x_1-x_2)^2+(y_1-y_2)^2+(z_1-z_2)^2)^{1/2}}  .
\end{equation}
The second integral can be calculated using the change of variables $s_x=x_1+x_2$, $s_y=y_1+y_2$, $t_x=x_1-x_2$, and $t_y=y_1-y_2$, giving
\begin{equation}
\frac{1}{4 \pi^2 L^4} \int dt_x dt_y e^{-(t_x^2+t_y^2)/2L^2} \int ds_x ds_y \frac{e^{-(s_x^2+s_y^2)/2L^2}}{(s_x^2+s_y^2+(z_1-z_2)^2)^{1/2}}=\frac{\sqrt{\pi}}{\sqrt{2}L}e^{(z_1-z_2)^2/2L^2}(1-\mathrm{erf}(|z_1-z_2|/\sqrt{2}L)) .
\end{equation}
Applying an additional change of variables, $s=z_1+z_2$ and $t=z_1-z_2$, we obtain
\begin{equation}
\frac{e^2}{4 \pi \epsilon_0 \epsilon_r} \frac{1}{8\sqrt{2\pi}l^2L} \int ds \ dt \ e^{-s^2/2l^2} e^{-t^2(L^2-l^2)/2L^2l^2} \left(\cos{(k_0t)}-\cos{(k_0s)}\right)^2 (1-\mathrm{erf}(|t|/\sqrt{2}L)).
\label{eq:intz}
\end{equation}
The following integrals are used to evaluate Eq.~(\ref{eq:intz}):
\begin{subequations}
\begin{equation}
\int du \ e^{-\kappa u^2} = \sqrt{\frac{\pi}{\kappa}}, \ \kappa>0
\label{eq:intz1}
\end{equation}
\begin{equation}
\int du \ e^{-\kappa u^2} \cos{(k_0 u)}= \sqrt{\frac{\pi}{\kappa}}e^{-k_0^2/4\kappa}, \ \kappa>0
\end{equation}
\begin{equation}
\int du \ e^{-\kappa u^2} \cos^2{(k_0 u)}= \frac{(e^{-k_0^2/\kappa}+1)\sqrt{\pi}}{2 \sqrt{\kappa}}, \ \kappa>0
\end{equation}
\begin{equation}
\int du \ e^{-\kappa u^2} \mathrm{erf(\tau |u|)}= 2\arctan{(\tau/\sqrt{\kappa})}/\sqrt{\pi \kappa}, \ \kappa>0, \ \tau>0
\end{equation}
\begin{equation}
\int du \ e^{-\kappa u^2} \cos^2{(k_0 u)} \ \mathrm{erf(\tau |u|)}= \int_0^{\infty} du \ e^{-\kappa u^2} \cos{(2 k_0 u)} \ \mathrm{erf(\tau u)}+\int_0^{\infty} du \ e^{-\kappa u^2} \ \mathrm{erf(\tau u)}, \ \kappa>0, \ \tau>0.
\label{eq:long_int}
\end{equation}
\end{subequations}
Here, the terms with factors $e^{-k_0^2/4\kappa}$ and $e^{-k_0^2/\kappa}$ are negligible. To calculate the first term on the right-hand side of Eq.~(\ref{eq:long_int}), we use integration by parts:
\begin{equation}
\begin{aligned}
\int_0^{\infty} du \ e^{-\kappa u^2} \cos{(2 k_0 u)} \ \mathrm{erf(\tau u)}&=\left.\frac{e^{-\kappa u^2} \ \mathrm{erf}(\tau u) \sin(2k_0u)}{2k_0} \right|_0^{\infty}-\frac{1}{2k_0}\int_0^{\infty} du \ \sin{(2k_0u)} \left( \frac{2\tau}{\sqrt{\pi}}e^{-(\kappa +\tau^2)u^2}-2\kappa u \ \mathrm{erf}(\tau u )\right) \\
&=0-\frac{\tau \ F(k_0/\sqrt{\kappa+\tau^2})}{k_0 \sqrt{\pi(\kappa+\tau^2)}} +\frac{\kappa}{k_0}\int_0^{\infty} du \ \sin{(2k_0u)} u \ \mathrm{erf}(\tau u),
\end{aligned}
\label{eq:dawson}
\end{equation}
\end{widetext}
where $F$ is the Dawson's integral~\cite[Eq.~7.2.5]{nist}. We can also use integration by parts to evaluate the remaining integral in Eq.~(\ref{eq:dawson}), which would yield terms proportional to the higher orders of $1/k_0$. These terms are also negligible. Combining everything, we obtain
\begin{multline}
\iinteg{1}{1}{1}{1} \approx \frac{e^2}{ 4 \pi \epsilon_0 \epsilon_r \sqrt{2\pi (L^2-l^2)}} \Biggl( \pi + \\
\frac{\sqrt{L^2-l^2}F(\sqrt{2} k_0l)}{\sqrt{2}k_0L^2}-2\arctan{\left(\frac{l}{\sqrt{L^2-l^2}}\right)} \Biggr).
\label{eq:m1111}
\end{multline}

\begin{figure}[b]
\includegraphics[width=2.5 in]{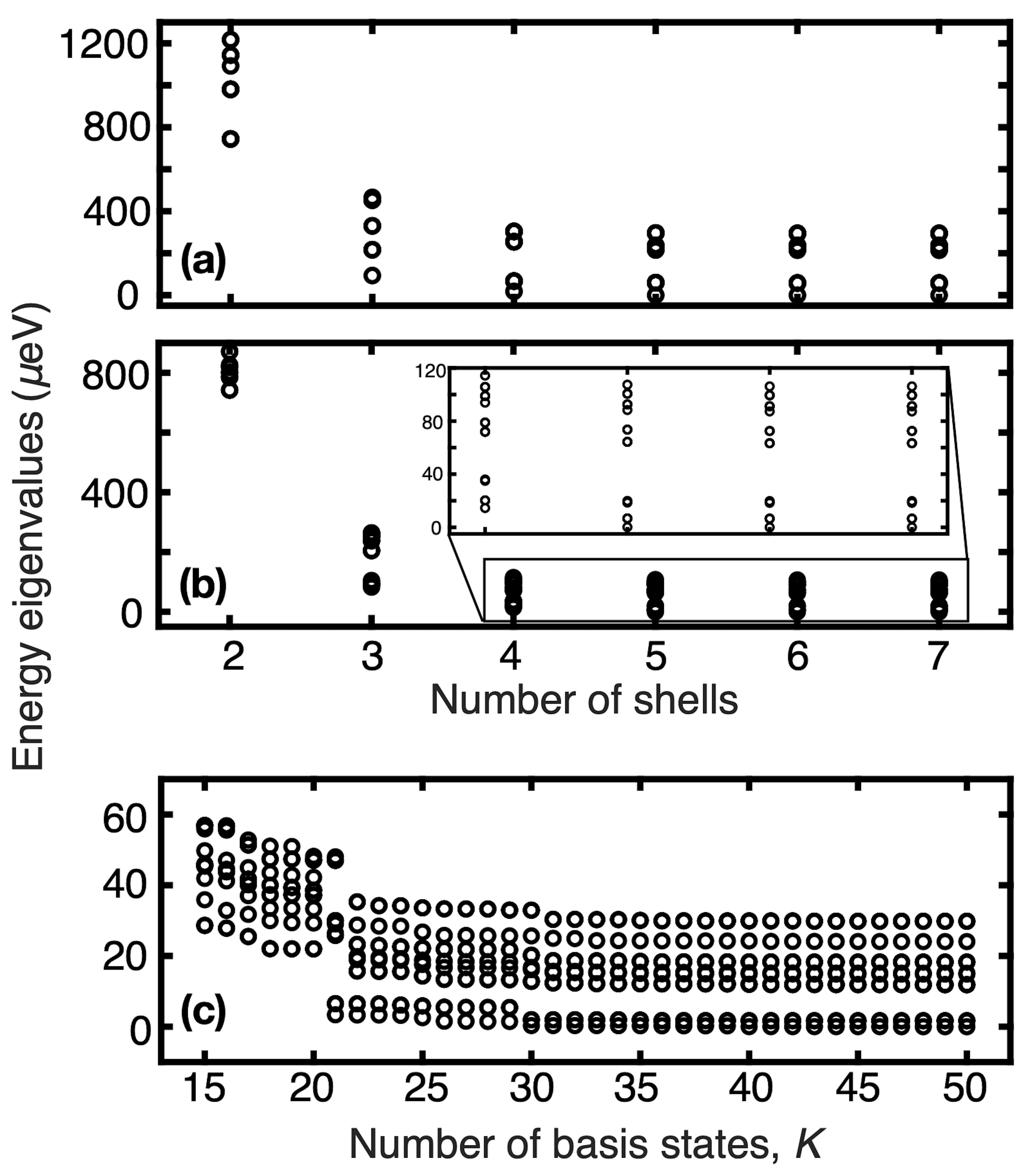}
\caption{Convergence properties of the FCI calculations. Data correspond to (a) the case with $\hbar \omega=0.4$~meV, in Fig.~\ref{fig:fig3L}(b), (b) the case with $\hbar \omega=0.5$~meV, $W=0.6D$, and $d=0$ in Fig.~\ref{fig:A4L}, and (c) the case with $\hbar \omega_y=0.8 \hbar \omega_x=0.24$~meV in Fig.~\ref{fig:fig6L}(b). Results for (a),(b) the lowest 20, or (c) the lowest 16 levels (including degeneracies), with respect to the ground state, calculated using up to (a),(b) 7 shells ($K=56$), or (c) 50 single-electron basis states. }
\label{fig:A5L}
\end{figure}

Similarly, $\iinteg{1}{1^\prime}{1}{1^\prime}$ can be calculated by replacing $\left(\cos{(k_0t)}-\cos{(k_0s)}\right)^2$ term in Eq.~(\ref{eq:intz}) by $\left(\cos{(2k_0t)}-\cos{(2k_0s)}\right)/2$. The only term that does not have an exponential suppression in this integral is a product of the integrals shown in Eqs.~(\ref{eq:intz1}) and(\ref{eq:dawson}), yielding
\begin{equation}
\iinteg{1}{1^\prime}{1}{1^\prime} \approx \frac{e^2}{ 4 \pi \epsilon_0 \epsilon_r} \frac{F(\sqrt{2} k_0l)}{2 \sqrt{\pi}k_0L^2}.
\label{eq:m1212}
\end{equation}

Some values of the integrals obtained using Eqs.~(\ref{eq:m1111}) and (\ref{eq:m1212}), along with the values obtained by using the numerical method summarized in Appendix~\ref{sec:app1}, for the wavefunction forms in Eq.~(\ref{eq:env_wfs}), are shown in Table~\ref{tab:integrals}. We find that $\iinteg{1}{1}{1}{1}$ is indeed much larger than $\iinteg{1}{1^\prime}{1}{1^\prime}$, and there is good agreement between the analytical and numerical methods. We finally note that the values in the third row of Table~\ref{tab:integrals} best correspond to the values in Fig.\ \ref{fig:A3L}. Here, we chose the $l$ value to yield a similar inverse participation ratio in $z$ as  in Fig.~\ref{fig:A3L}, in which $\iinteg{1}{1}{1}{1}=7.9$~meV and $\iinteg{1}{1^\prime}{1}{1^\prime}=0.2$~$\mu$eV. 

\section{\label{sec:app3} Additional details regarding Fig.~\ref{fig:fig5L}}
In Fig.~\ref{fig:A4L}(a), we show the ground state electron densities, as a function of $x$, for the data in Fig.~\ref{fig:fig5L}. Here, the blue lines denote the positions of the atomic steps at the interface. Figure~\ref{fig:A4L}(b) shows the corresponding $E_{\mathrm{ST}}$ and $E_{\mathrm{val}}$ values, together with the interface profiles considered.

\section{\label{sec:app4} FCI convergence}
This Appendix reports the convergence properties of the FCI calculations. In Figs.~\ref{fig:A5L}(a) and \ref{fig:A5L}(b), we plot the energy eigenvalues of the most-weakly-confined circular dots, as a function of the number of shells included in the single-electron basis set. Due to the two-fold multiplicity of the valleys, $n$ shells correspond to $K=n(n+1)$ states, excluding spin. In the circular dot calculations we used 6 shells, as this number is observed to be sufficient for convergence in the most challenging cases. In Fig.~\ref{fig:A5L}(c), we plot the energy eigenvalues of the most-weakly-confined elliptical dots as a function of the number single-electron states included basis, as there is no well-defined shell structure appropriate for all cases in Fig.~\ref{fig:fig6L}. We used $K=40$ in these calculations.

\bibliography{text}

\end{document}